\documentclass[authoryear,preprint,review,12pt]{elsarticle}

\usepackage{amssymb}
\usepackage{amsmath}
\usepackage{hyperref}
\usepackage{booktabs}
\usepackage{adjustbox} 
\usepackage{multirow}
\usepackage{tabularx}
\usepackage{longtable}
\usepackage{booktabs}
\usepackage{ragged2e} 
\usepackage{xltabular}

\journal{Chemical Engineering Science}

\begin{document}

\begin{frontmatter}



\title{A comprehensive approach to incorporating intermolecular dispersion into the openCOSMO-RS model. Part 1: Halocarbons}


\author[inst1,inst11]{Daria Grigorash}
\affiliation[inst1]{organization={Department of Chemistry, Technical University of Denmark},
            city={Kgs. Lyngby},
            postcode={2800}, 
            country={Denmark}}
\affiliation[inst11]{organization={Center for Energy Resources Engineering, Technical University of Denmark},
            city={Kgs. Lyngby},
            postcode={2800}, 
            country={Denmark}}
            
\author[inst2]{Simon Müller}
\affiliation[inst2]{organization={Institute of Thermal Separation Processes, Hamburg University of Technology},
            city={Hamburg},
            postcode={21073}, 
            country={Germany}}

\author[inst3]{Patrice Paricaud}
\affiliation[inst3]{organization={UCP, ENSTA Paris, Institut Polytechnique de Paris},
            city={Palaiseau},
            postcode={91762}, 
            country={France}}
            
\author[inst1,inst11]{Erling H. Stenby}
\author[inst2]{Irina Smirnova}
\author[inst1,inst11]{Wei Yan\corref{corresponding author}}
\cortext[corresponding author]{Corresponding author at: Department of Chemistry, Technical University of Denmark, 2800 Kgs. Lyngby, Denmark.\ead{weya@kemi.dtu.dk}}
\begin{abstract}
The COSMO-RS (Conductor-like Screening Model for Real Solvents) is a predictive thermodynamic model that has found diverse applications in various domains like chemical engineering, environmental chemistry, nanotechnology, material science, and biotechnology. Its core concept involves calculating the screening charge density on the surface of each molecule and letting these surface patches interact with each other to calculate thermodynamic properties. In this study, we aim to enhance the performance of the open-source implementation openCOSMO-RS by incorporating dispersive interactions between the paired segments. Several parametrizations were systematically evaluated through the extensive regression analysis using a comprehensive database of Vapor-Liquid Equilibrium (VLE), Liquid-Liquid Equilibrium (LLE) and Infinite Dilution Activity Coefficients (IDACs). Furthermore, the influence of different combinatorial terms on the model performance was investigated. Our findings indicate that incorporating dispersive interactions significantly improves the accuracy of phase equilibrium predictions for halocarbons and refrigerant mixtures.
\end{abstract}


\begin{highlights}
\item The inclusion of the dispersion contribution enhanced openCOSMO-RS' ability to predict the phase equilibrium in halocarbons and refrigerants.
\item The model for halocarbons was developed by performing an extensive analysis of various parametrization approaches, differing in combinatorial terms and dispersion parameters. 
\item The developed model generally outperforms existing open-source implementations of COSMO-RS for the investigated halocarbon systems.
\end{highlights}

\begin{keyword}
COSMO-RS \sep Dispersion \sep Parameterization \sep Refrigerants 
\end{keyword}
\end{frontmatter}


\section{Introduction}
\label{sec:sample1}


Over the last two decades, the developments in computational chemistry methods have laid the foundation for a new class of predictive thermodynamic models originally based on the COSMO (COnductor-like Screening Model) \citep{Klamt1993COSMO:Gradient} quantum chemical (QC) method. It is one of the dielectric continuum solvation models that is used to determine the dielectric screening effects on molecules in solvents. The pioneering work by \citet{Klamt1995Conductor-likeMolecules} on integrating COSMO into a thermodynamic model created COSMO-RS. Since then, several implementations based on the COSMO-RS model, including COSMO-SAC \citep{Lin2002AModel}, COSMO-RS(ol) \citep{Grensemann2005PerformanceMethods} and \citet{JCOSMO} have been proposed. Additionally, the concept of segment-segment interactions appeared in the group-contribution method F-SAC \citep{Soares2013Functional-segmentFormulation}. However, in F-SAC, regression of experimental thermodynamic data replaced QC calculations. 

COSMO-RS is a predictive model that allows the estimation of thermodynamic properties even when no experimental data is available. Since its creation, COSMO-RS and its implementations have been assessed and applied for the prediction of solvation Gibbs free energies \citep{Saidi2020PredictionsApproaches}, VLE and IDACs across diverse chemical families of compounds \citep{Fingerhut2017ComprehensiveEquilibria}, and various thermodynamic properties in systems such as ionic liquids \citep{Jiriste2022PredictingCOSMO-RS, Lee2017PredictionModel,Diedenhofen2010COSMO-RSReview}, pharmaceuticals \citep{Klajmon2022PurelyCOSMO-RS, Rodriguez-Llorente2023ExtractionStudies,Loschen2016ComputationalSolvates,Klamt2002PredictionCOSMO-RS}, polymer solutions \citep{Silva2023TheBehavior,Loschen2014PredictionCOSMO-RS}, electrolytes \citep{Muller2019EvaluationIons, Muller2020CalculationSystems, GonzalezdeCastilla2021OnCoefficients} and refrigerants \citep{Mambo-Lomba2021PredictionsApproaches}. This study focuses on refrigerant systems, which are vital for industrial processes, especially in optimizing refrigerant production and operating energy systems like heat pumps. Understanding the phase behavior of refrigerant blends, including azeotropes and VLE, is critical for selecting their optimal combinations. 

Within the COSMO-RS framework, molecules are fragmented into interacting segments, primarily considering electrostatic interactions and hydrogen bonding. These interactions are characterized by a set of descriptors or properties of these segments, with the surface charge density being the most important among them. In the early versions of COSMO-RS, dispersive intermolecular interactions were not accounted for in fluid mixtures. These interactions are caused by attractive forces between instantaneous dipoles \citep{Prausnitz1999MolecularEquilibria}. Since these dipoles are not permanent, their interactions could not be captured with classic electrostatics \citet{Stone2013TheForces}. Nevertheless, dispersive interactions have already been considered in several implementations of COSMO-RS. For instance, COSMO-SAC-dsp by \citet{Hsieh2014ConsideringBehavior} introduces a one-constant Margules term to activity coefficients. The constant is determined from atomic dispersion parameters which are general for the model. As previously mentioned by \citet{Krooshof2019DispersionStateb}, this treatment greatly simplifies the dispersion phenomenon, since using the Margules model implies that dispersion comes only from similar-sized interactions. Moreover, in mixtures of species that belong to one aliphatic hydrocarbon homologous series, dispersion is neglected, since the contribution of hydrogen (H-atom) in such compounds equals to zero. Therefore, other equations for the dispersive contribution to the activity coefficient were proposed in works by \citet{Krooshof2019DispersionState, Krooshof2019DispersionStateb}. However, to our knowledge, none of those is implemented in the open-source software. \citet{Mambo-Lomba2021PredictionsApproaches} modified the atomic dispersion parameter of fluorine (F-atom) in the COSMO-SAC-dsp model, which significantly improved VLE predictions of the studied refrigerants. For high-pressure systems, the MHV1 \(EoS/G^E\) mixing rule approach was used to combine the modified COSMO-SAC-dsp with the Peng-Robinson equation of state. In the F-SAC model, dispersion is addressed via general group interaction parameters \citep{Flores2016IncludingModelb}. The group - group interactions in F-SAC are analogous to segment - segment interactions in COSMO-RS. Therefore, the integration of dispersive interactions in F-SAC is conceptually different from COSMO-SAC-dsp. Furthermore, in the recent versions of COSMOtherm software \citep{COSMOtherm}, a dispersion term is also introduced for fluid mixtures, using the matrix of pairwise element-element specific van der Waals parameters \citep{Klamt2007PredictionCOSMOtherm}. 

The major challenge in modeling the phase equilibrium of halocarbons as part of refrigerants is the substantial difference in polarizabilities of halogens, e.g., fluorine  having an extremely low polarizability in contrast to iodine having a very high one. Therefore, there is a clear need for special consideration of the dispersive contribution regarding this type of systems.

In the present study, we aim to develop the dispersion contribution to the segment-segment interactions, currently absent in the open-source implementation of COSMO-RS: openCOSMO-RS by \citet{Gerlach2022AnDescriptors}. To outline the main concepts of COSMO-RS, its major equations are presented in the Methods \ref{sec:Methods} section: residual contribution, including electrostatic misfit and hydrogen bonding in section \ref{sec:COSMORS}, the newly-implemented dispersive contribution in section \ref{sec:COSMORS_disp}, and various combinatorial terms in section \ref{sec:combinatorial_terms_methods}. The workflow used for the QC calculations and regression procedure are explained in sections \ref{sec:COSMO_calculations} and \ref{sec:regression_methods}, respectively. The database introduced in section \ref{sec:database} was used for the regressions discussed in section \ref{sec:Parameterizations_discussion}. Subsequently, we evaluated the impact of the combinatorial term on the IDAC of hydrocarbons in section \ref{sec:combinatorial_term_discussion}. Finally, in section \ref{sec:VLE_and_LLE_predictions}, we examined the VLE and LLE predictions by the modified models and compared them with existing open-source COSMO-RS implementations.

\section{Methods}
\label{sec:Methods}
\subsection{COSMO-RS}
As most of the existing activity coefficient models, COSMO-RS includes both residual \ref{sec:COSMORS}, \ref{sec:COSMORS_disp} and combinatorial \ref{sec:combinatorial_terms_methods} terms:
\begin{equation}
\ln(\gamma_i) = \ln(\gamma_i^{\text{res}}) + \ln(\gamma_i^{\text{comb}}).
\end{equation}
\subsubsection{Residual contribution: electrostatic misfit and hydrogen bonding}
\label{sec:COSMORS}

The detailed equations used in openCOSMO-RS are reported in \citet{Klamt1998RefinementCOSMO-RS}, \citet{Klamt2000COSMO-RS:Liquids} and \citet{Gerlach2022AnDescriptors}.
Following the COSMO-RS approach, molecules are represented by a set of surface patches or segments, each characterized by a group of descriptors. One of the primary descriptors is the surface charge density (\(\sigma\)). It is determined using the COSMO or similar C-PCM (Conductor-like Polarizable Continuum Model) QC method. Continuum solvation models represent each solvent as a dielectric continuum with a dielectric permittivity \(\epsilon\). The solute molecule is then embedded into this dielectric continuum, which, in the case of COSMO, is modeled as a perfect conductor with \(\epsilon=\infty\) \citep{KLAMT200511}. In response, screening charge is generated on the surface of the molecular cavity, constructed using element-specific radii. While \(\sigma\) serves as a major descriptor, the openCOSMO-RS implementation used in this study also incorporated additional descriptors and simplified the procedure for integrating new ones into the model if needed. 

In COSMO-RS, thermodynamic properties are related to the interactions between molecular segments:\\
\begin{equation}
\mu_s(\sigma_i) = -kT\ln(\int d\sigma_jp_s(\sigma_j)\exp{[(-E_{ij}+\mu_s(\sigma_j))/kT]},   
\end{equation}
where \(\sigma_i\) and \(\sigma_j\), \(\mu_s(\sigma_i)\) and \(\mu_s(\sigma_j)\) are the surface screening charge densities and chemical potentials of segments \textit{i} and  \textit{j}, respectively. \textit{k} is Boltzmann's constant, and \textit{T} is the temperature. \(p_s(\sigma_j)\) or the \(\sigma\)-profile defines the probability of finding the segment with charge density \(\sigma_j\) on the surface of the molecule. It is formed by clustering segments onto a predetermined grid of selected values of \(\sigma_j\) and projecting them onto a histogram. Before clustering segments into the \(\sigma\)-profile, they are averaged over a region of radius \(r_{av}\). The areas of the segments are an output from the currently used COSMO/C-PCM implementation of the underlying QC package. However, these areas might significantly differ from each other. Therefore, averaging over a region of radius \(r_{av}\) is performed to spread the charge and to reduce numerical noise. Finally, \(E_{ij}\) represents the interaction free energy between segments \textit{i} and \textit{j}. 

In the early versions of COSMO-RS \citep{Klamt1998RefinementCOSMO-RS,Klamt2000COSMO-RS:Liquids}, two contributions to the interaction free energy were considered for modeling the condensed phase equilibrium: the repulsive misfit free energy \(E_{mf}(\sigma_i,\sigma_j)\) and the attractive hydrogen bonding energy \(E_{hb}(\sigma_i,\sigma_j)\). The former captures electrostatic interactions between the segments, while the latter represents an additional attractive contribution arising from hydrogen bond formation. 
The electrostatic misfit free energy is calculated as:
\begin{equation}
E_{mf}(\sigma_i,\sigma_j) = 0.5a_{eff}\alpha_{mf}[(\sigma_i+\sigma_j)^2+f_{corr}(\sigma_i+\sigma_j)(\sigma_i^\perp+\sigma_j^\perp)],
\end{equation}
involving both the screening charge density \(\sigma\) and the correlation screening charge density \(\sigma^\perp\) to account for the effect of the segment surroundings. \(f_{corr}\) is the correlation correction factor adjusted to the dielectric energy data \citep{Klamt1998RefinementCOSMO-RS}. The general COSMO-RS parameters \(\alpha_{mf}\) and \(a_{eff}\) are the misfit prefactor and the effective contact area of a segment, respectively.
The hydrogen bond contribution, assuming that segment \textit{i} as a donor and \textit{j} is an acceptor (\(\sigma_i < \sigma_j\)), is expressed as:
\begin{equation}
 E_{hb}(\sigma_i,\sigma_j) = a_{eff}c_{hb}(T)[\min(0;\sigma_i+\sigma_{hb})\max(0;\sigma_j-\sigma_{hb})].
\end{equation}
Due to this functional form, a non-zero hydrogen bond contribution would be observed only for segments with opposing signs of \(\sigma\), the absolute value of which exceeds the threshold value of \(\sigma_{hb}\). This ensures that this effect is considered only for highly polar segments.
To account for the temperature dependence of hydrogen bond formation, the following temperature-dependent coefficient \(c_{hb}(T)\) is introduced:
\begin{equation}
c_{hb}(T) = c_{hb}\max(0;1-c_{hb}^T+c_{hb}^T\cdot298.15/T)
\end{equation}

At this stage, seven general COSMO-RS parameters are considered:  \(r_{av}\), \(\alpha_{mf}\), \(a_{eff}\), \(f_{corr}\), \( c_{hb}\), \(c_{hb}^T\) and \(\sigma_{hb}\). Since in this work we focus only on halocarbons, the hydrogen bonding contribution is irrelevant. 
\subsubsection{Residual contribution: dispersive interactions}
\label{sec:COSMORS_disp}
The interaction energy between two segments \(E_{ij}\) does not initially include the dispersion contribution, which is arguably negligible for highly polar mixtures but crucial for systems of low polarity such as halocarbons. In this study, we incorporated the dispersion energy of interactions between segments \textit{i} and \textit{j} as:
\begin{equation}
 E_{vdW} = (1-k_{ij})a_{eff}\tau_i^{vdW}\tau_j^{vdW}, \label{eq:Evdw}
\end{equation}
resulting in the final equation for the segment-segment interaction free energy:
\begin{equation}
 E_{ij} = E_{mf}(\sigma_i,\sigma_j) + E_{hb}(\sigma_i,\sigma_j) - E_{vdW}. 
\end{equation}
Here \(\tau_i^{vdW}\) and \(\tau_j^{vdW}\) are the dispersion coefficients of segments \textit{i} and \textit{j}, respectively. These parameters are unique for each atom. When contacting segments belong to different atoms, the non-zero coefficient \(k_{ij}\) may be used to enhance the accuracy of predictions. However, this significantly increases the number of adjustable parameters. 

The functional form in Eq.\ref{eq:Evdw} was inspired by the London attractive potential \citep{London1937TheForces}:
\begin{equation}
 \phi^{\text{London}}_{ij} = -\frac{3}{2}\frac{\alpha_{i}\alpha_{j}}{r_{ij}^6}\cdot\frac{h\nu_ih\nu_j}{h(\nu_i+\nu_j)}, \label{eq:London}
\end{equation}
which quantifies the attractive potential between two spherically symmetrical systems with polarizabilities \(\alpha_{i}\) and \(\alpha_{j}\) separated by the distance \(r_{ij}\) from each other. Additionally, Eq.\ref{eq:London} demonstrates the influence of zero-point vibrations with their frequencies \(\nu_i\) and \(\nu_j\). 

Within the COSMO-RS framework, direct consideration of the London potential is not feasible due to central assumptions and the two-dimensional nature of the statistical mechanics involved \citep{KLAMT200559}. However, the functional form of the dispersion parameters in Eq. \ref{eq:Evdw} resembles the dependence of the London potential on atomic polarizabilities in Eq.\ref{eq:London}. 
\subsubsection{Combinatorial contribution}
\label{sec:combinatorial_terms_methods}
For the combinatorial contribution, we consider the following widely used terms: the Staverman–Guggenheim term \citep{Staverman1950TheFormulae} Eq.\ref{eq:SG_term}, the Flory-Huggins term  \citep{Flory1942ThemodynamicsSolutions} Eq.\ref{eq:FH_term} and the Flory-Huggins term modified by \citet{Elbro1990ASolutions}. 
The equation for the Staverman–Guggenheim (SG) term is:
\begin{equation}
\ln(\gamma_i^{\text{comb}}) = \ln(\frac{\phi_i}{x_i}) + 1 - \frac{\phi_i}{x_i} - 0.5z \frac{A_i}{A_{std}}(\ln(\frac{\phi_i}{\theta_i}) + 1 -\frac{\phi_i}{\theta_i}),\label{eq:SG_term}
\end{equation}
with the volume fractions \(\phi_i\) and the surface fractions \(\theta_i\) calculated as:
\begin{equation}
\frac{\phi_i}{x_i} = \frac{V_i}{\sum_{j}x_jV_j }, 
\end{equation}
\begin{equation}
\frac{\theta_i}{x_i} = \frac{A_i}{\sum_{j}x_jA_j }.
\end{equation}
In the equations above, \(V_i\) represents the hard-core volume, \(A_i\) is the molecular cavity surface area determined using the QC calculations.
For the SG term, the coordination number \textit{z} was set to 10, and \(A_{std}\) was treated as an adjustable parameter. 
The Flory-Huggins (FH) term is expressed as:
\begin{equation}
\ln(\gamma_i^{\text{comb}}) = \ln(\frac{\phi_i}{x_i}) + 1 - \frac{\phi_i}{x_i}. \label{eq:FH_term}
\end{equation}
Elbro et.al proposed to use 'free' volume \(V'_i\) instead of \(V_i\) in Eq.\ref{eq:FH_term}, calculated as:
\begin{equation}
V'_i = V_i - V_i^m,
\end{equation}
 where \(V_i^m\) is the molar volume determined from the experimental densities of pure compounds. 
\subsection{QC calculations}
\label{sec:COSMO_calculations}
For QC calculations, the open-source workflow described in detail in \cite{Gerlach2022AnDescriptors} was used. Following this methodology, all calculations were performed in ORCA 5.0.3 \citep{Neese2012TheSystem, Neese2020ThePackage}. Initially, a set of conformers was generated using RDKit Python library \citep{RDKit2023.03.3, Ebejer2012FreelyThey}, followed by ALPB \citep{Bannwarth2021ExtendedMethods, Ehlert2021RobustMethods} geometry optimizations in water using GFN2-xTB \citep{xTB} calculations from within ORCA. Then the conformers were filtered and their geometry was optimized with C-PCM at DFT/BP86/def2-TZVP(-f) level. Subsequently, the conformer with the lowest energy was selected and its geometry was optimized at DFT/BP86/def2-TZVP. Finally, C-PCM calculations were performed at DFT/BP86/def2-TZVPD level for the most stable conformer. It is worth mentioning that although C-PCM is not identical to the originally used COSMO method, these two methods are essentially the same for a solvent with a dielectric constant \(\epsilon=\infty\). An archive with examples of ORCA files can be found in Supplementary Materials. The entire procedure is automated in the \href{https://github.com/TUHH-TVT/openCOSMO-RS_conformer_pipeline}{openCOSMO-RS conformer generator pipeline} code, and the tools involved in all these steps are freely accessible for \textit{academia}.
\subsection{Regression procedure}
\label{sec:regression_methods}
To optimize COSMO-RS parameters, various types of phase equilibrium data were used: IDAC, partition coefficients of component \textit{i} between water and organic liquid (\(K_i^{\text{org/w}}\)), VLE and LLE.
For binary VLE, we calculated experimental activity coefficients using saturated vapor pressures of pure components either directly from experimental data, or if unavailable, calculated from the correlations provided by DIPPR \citep{DIPPR} , NIST \citep{Burgess2024NIST69} or \citep{Yaws2015VaporCompounds}. These experimental activity coefficients \( \ln(\gamma_i^{exp})\) were then used to minimize the following objective function:
\begin{equation}
OF = \sum_{s}^{Ns} w_s\sum_{i}^{Nc} (\ln(\gamma_i^{calc}) - \ln(\gamma_i^{exp}))^2, \label{eq:OF_VLE}
\end{equation}
where \(N_s\) is the number of datasets, and  \(N_c\) is the number of components for which data is available. All weights \(w_s\) were set to unity.
The same objective function Eq. \ref{eq:OF_VLE} was used to regress IDAC data.

The distribution coefficient \(K_i\) of component \textit{i} between phases \(\alpha\) and \(\beta\) is calculated as:
\begin{equation}
K_i^{\text{exp}} \equiv \frac{x_i^{\beta,\text{exp}}}{x_i^{\alpha, \text{exp}}} = \frac{\gamma_i^{\alpha,\text{exp}}}{\gamma_i^{\beta, \text{exp}}}.\label{eq:Ki}
\end{equation}
Here, the ratio of mole fractions in separated phases on the left-hand side of this equation was determined from the LLE experimental data, while the ratio of activity coefficients was calculated with openCOSMO-RS at experimental concentration. We utilized the following objective function for the LLE data regression:
\begin{equation}
OF = \sum_{s}^{Ns} w_s\sum_{i}^{Nc} ( \ln(K_i^{\text{calc}}) - \ln(K_i^{\text{exp}}) )^2, \label{eq:OF_LLE}
\end{equation}
which was also used for partition coefficients \(K_i^{\text{org/w}}\).

\section{Results and discussion}
\label{sec:Results and discussion}

\subsection{Database}
\label{sec:database}
To optimize the dispersion parameters for the openCOSMO-RS model, we focused on three types of hydrocarbon and halocarbon thermodynamic data. Table \ref{tab:database} provides an overview of the collected datasets, including their types, the number of datasets, and the number of data points used for regressions. More details on the sources of the datasets are reported in Table S1 of the Supplementary Materials. 

\begin{table}[htbp]
\caption{An overview of the collected experimental data on hydrocarbons and halocarbons.}
\label{tab:database}
\centering
\begin{tabular}{l l l}
\hline
\textbf{Type} & \textbf{Data points} & \textbf{Datasets} \\
\hline
IDAC & 310 & 224 \\
LLE & 1534 & 85 \\
VLE & 10104 & 397 \\
\hline
\end{tabular}
\end{table}

The majority of the VLE datasets are from the Korean Database \citep{CHERIC2024} and from the database by \citep{Jaubert2020BenchmarkAccuracy}. While the former did not report any thermodynamic consistency analysis of the data, the latter was specifically developed for benchmarking thermodynamic models, ensuring a quality assessment of the data. To further assure data quality, we performed the van Ness thermodynamic consistency test \citep{VanNess1995ThermodynamicsData} on the VLE data. Additionally, all datasets were plotted in \(\ln(\frac{\gamma_1}{\gamma_2}) - x_1\) coordinates to visually verify a reasonable trend. Approximately half of the initially collected datasets were eliminated due to inconsistencies.

For LLE datasets, a significant fraction of experimental studies carried out the cloud point measurements, which provide only one end of the tie-line. Thus, we correlated these data with the UNIQUAC thermodynamic model \citep{Abrams1975StatisticalSystems,Anderson1978ApplicationEquilibria} and extrapolated the missing tie-lines. 
\begin{figure}
    \centering
    \includegraphics[width=0.45\linewidth]{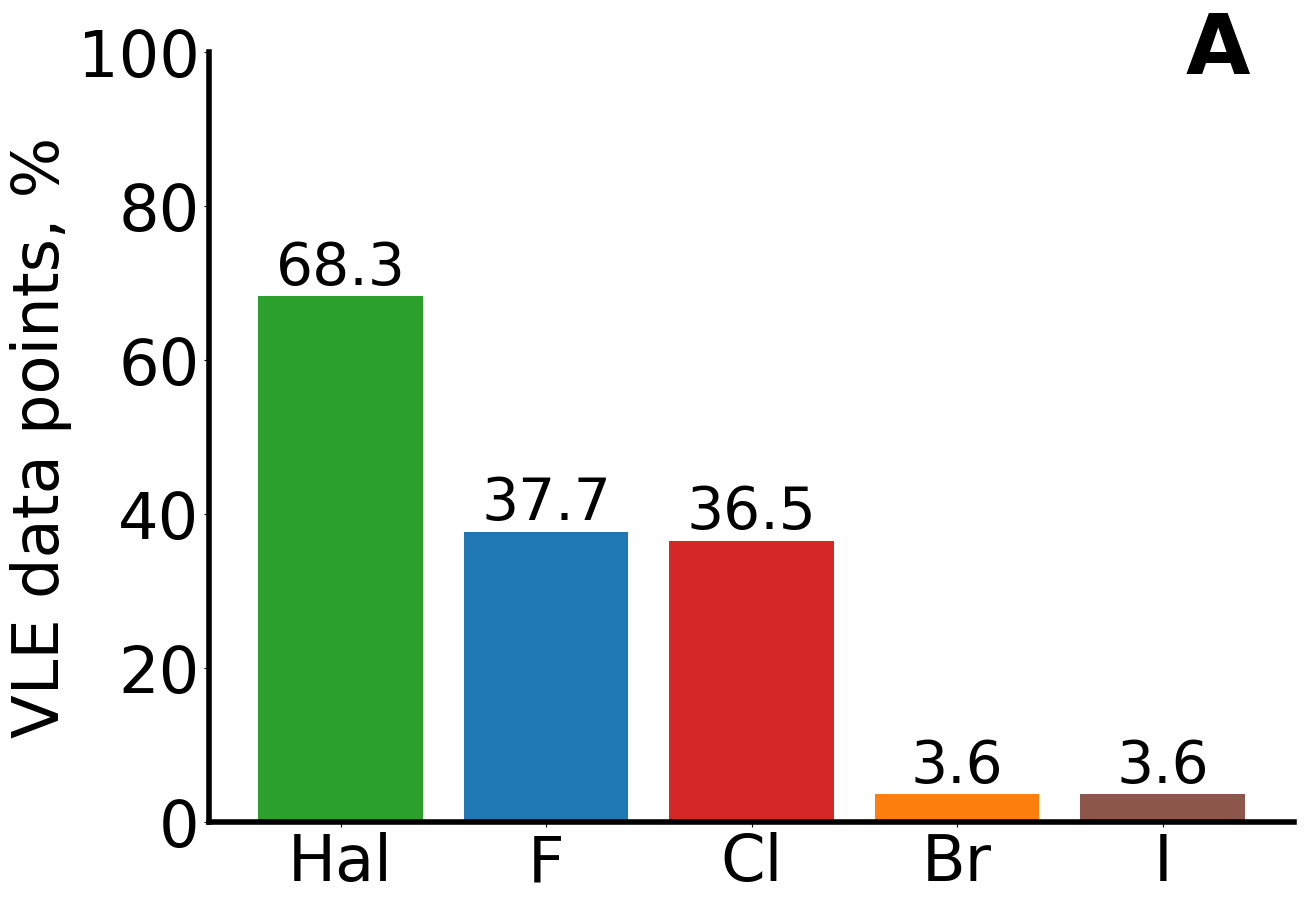}
    \includegraphics[width=0.45\linewidth]{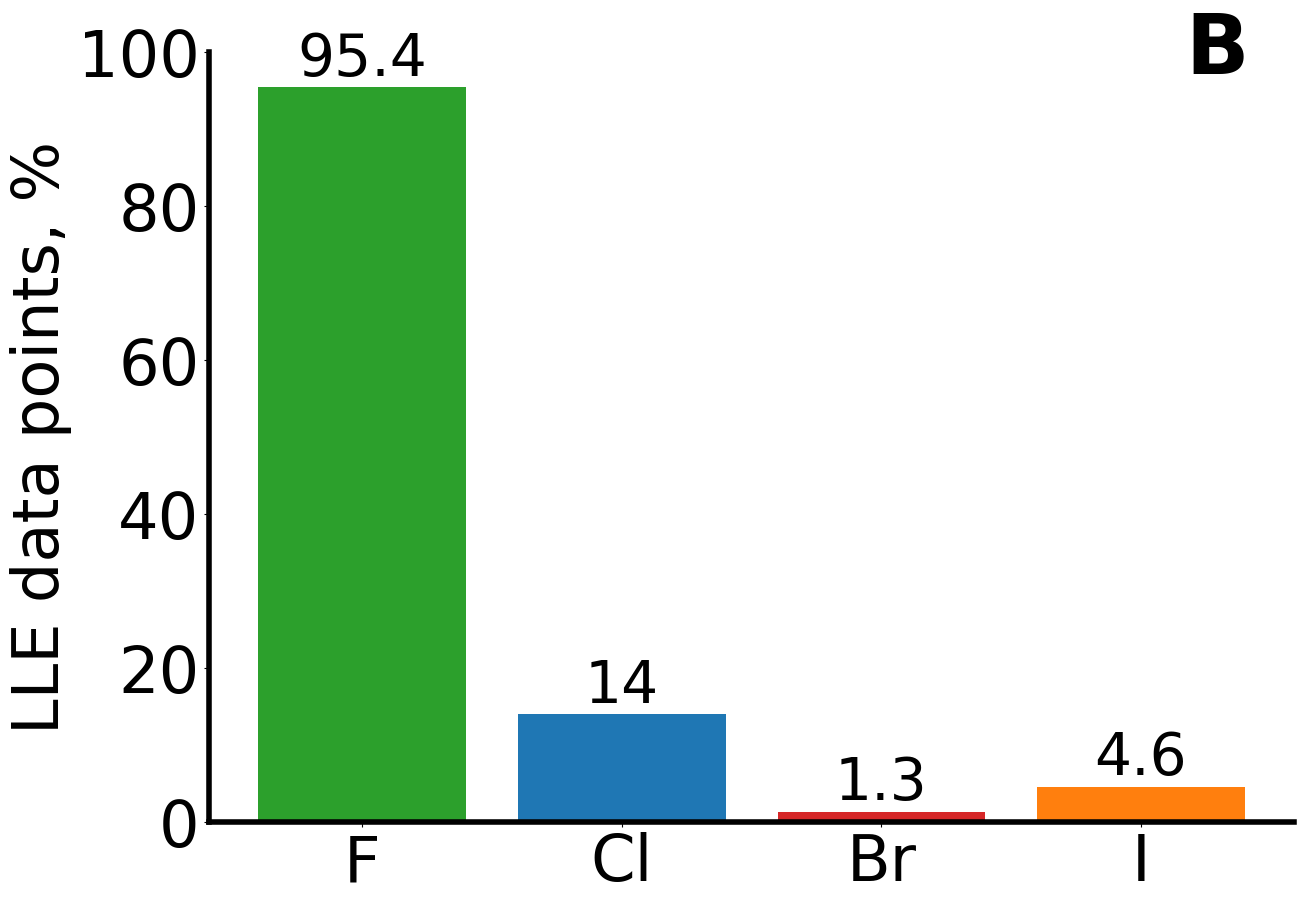}
    \caption{The distribution of halogen-atoms (Hal) in the VLE (\textbf{A}) and LLE (\textbf{B}) data.}
    \label{fig:VLE_LLE_bar}
\end{figure}
\begin{figure}
    \centering
    \includegraphics[width=0.5\linewidth]{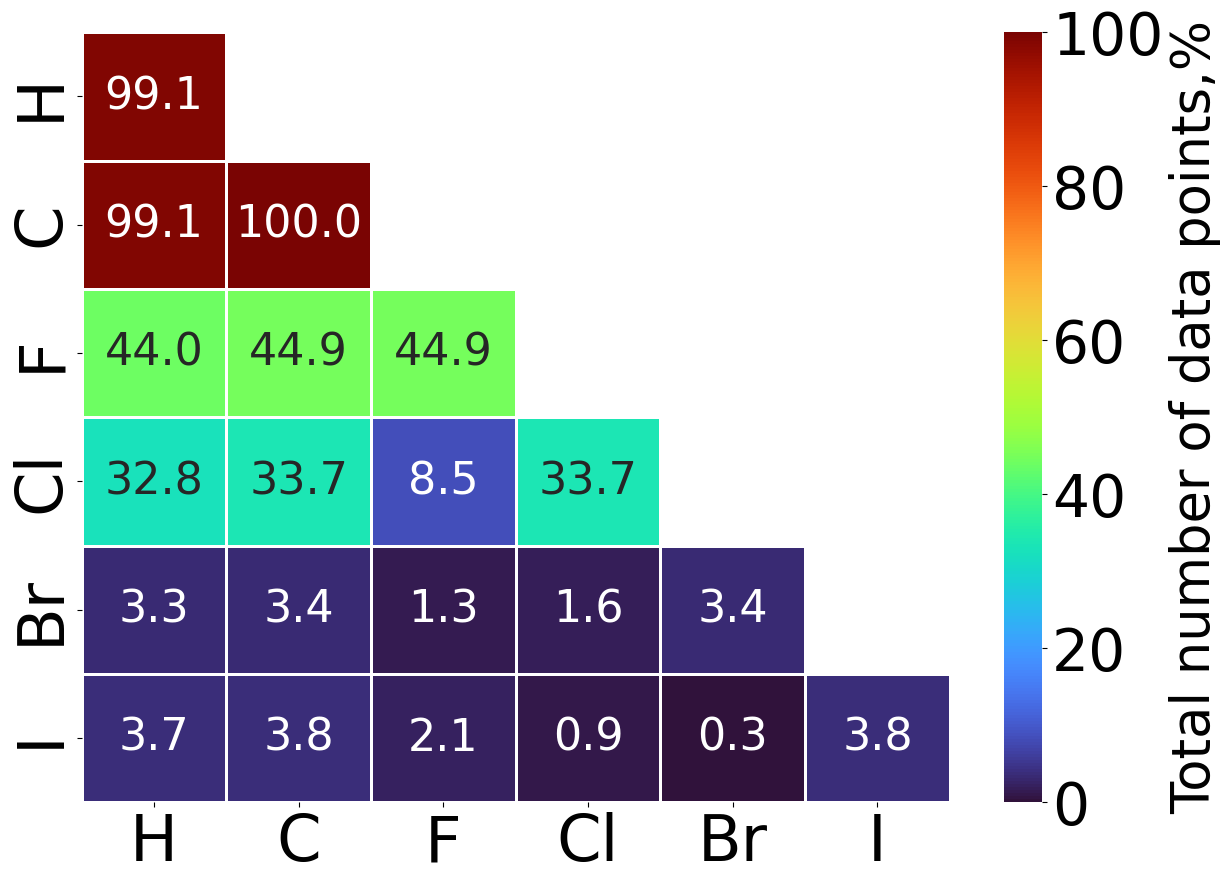}
    \caption{The frequency of occurrence of atom-atom pairs across the compiled database.}
    \label{fig:cross-atom}
\end{figure}
As shown in Figure \ref{fig:VLE_LLE_bar}, the VLE data exhibit a prevalence of both fluorinated and chlorinated compounds, while the LLE data primarily feature florinated compounds. Although the experimental data for these compounds are more abundant than those for brominated and iodinated hydrocarbons, it is important to include all species to cover all possible halogen-halogen combinations. The frequency of atom-atom pair occurrence in the collected data is illustrated in Figure \ref{fig:cross-atom}. For instance, the binary mixture of n-pentane and dichloromethane comprises various atom-atom contacts, including H-C, C-C, C-Cl, H-Cl. Each data point related to this mixture is assigned all the listed pairs to cover the possible segment-segment interactions with atomic number as a descriptor. To illustrate this, we calculated the percentage of an atom-atom pair in the total number of data points. It is evident from Figure \ref{fig:cross-atom} that the data for mixtures containing both Br and I atoms, or Cl and I atoms, is particularly scarce.

\subsection{Parametrization procedures}
\label{sec:Parameterizations_discussion}

To incorporate the dispersion term into openCOSMO-RS, we introduced several atom-specific and cross-atom parameters (Eq.\ref{eq:Evdw}). The parametrization approaches considered in this study, along with their corresponding abbreviations for further reference, are presented in Table \ref{tab:parameters}.

\begin{table}[htbp]
    \centering
    \caption{The characteristics and abbreviations of the parametrizations.}
    \label{tab:parameters}
    \begin{adjustbox}{width=\textwidth} 
        \begin{tabular}{lcccc}
            \toprule
            Combinatorial term & \multicolumn{2}{c}{Atom-specific parameters} & \multicolumn{2}{c}{Atom-specific parameters + cross-atom coefficients} \\
            \cmidrule(lr){2-3} \cmidrule(lr){4-5}
            & 6 parameters & 7 parameters & 6 parameters & 7 parameters \\
            \midrule
            Staverman–Guggenheim & SG\_6 & SG\_7 & SG\_6\_cross & SG\_7\_cross \\
            Flory-Huggins & FH\_6 & FH\_7 & FH\_6\_cross & FH\_7\_cross \\
            Elbro & Elbro\_6 & Elbro\_7 & Elbro\_6\_cross & Elbro\_7\_cross \\
            \bottomrule
        \end{tabular}
    \end{adjustbox}
\end{table}

At this stage, our focus is on hydrocarbons and halocarbons, however, the approach will be extended to other types of compounds in the future. Therefore, currently, the initial parameter set corresponds to C, H, F, Cl, Br, and I atoms, resulting in a total of 6 parameters. It is worth mentioning that the polarizability of C-atom vary with its hybridization. Hence, we also used two parameters to describe C\((sp^3)\) and C\((sp^2)\) atoms, leading to 7 parameters in total. The inclusion of cross-interaction parameters \(k_{ij}\) (Eq.\ref{eq:Evdw}) expands the total number of parameters to 28 when accounting for differences in C-atom hybridization, and 21 if not. Furthermore, all these parametrization schemes were incorporated with three combinatorial terms (SG, FH, Elbro \ref{sec:combinatorial_terms_methods}), resulting in 12 possible combinations.
From this point onward, the ORCA-B parametrization with the SG term from \citet{Gerlach2022AnDescriptors} is considered as the benchmark without the dispersion term, and is denoted as 'no dispersion' in subsequent figures. 
Tables \ref{tab:param_deviation} and \ref{tab:param_deviation_cross} summarize all the parameters regressed in this study. Table \ref{tab:param_deviation} covers parametrizations with only 6 or 7 atom-specific dispersion parameters. For SG\_6 and SG\_7, the general model parameters (\(a_{eff}\), \(r_{av}\), \(\alpha_{mf}\), \(f_{corr}\), \(c_{hb}\), \(c_{hb}^T\), \(\sigma_{hb}\), and \(A_{std}\)) were fixed to the corresponding values of ORCA-B parametrization. For the FH term, this set of parameters was regressed to the \(K_i^{\text{org/w}}\) and IDAC data using the same database as for SG\_6 and SG\_7 reported in \citet{Gerlach2022AnDescriptors}. With the Elbro term, we followed a different approach by regressing both general and dispersion parameters simultaneously only to the data in Table \ref{tab:database} due to poor performance of the Ebro term for the highly polar and aqueous systems, which are abundant in IDAC and \(K_i^{\text{org/w}}\) data collection. Consequently, the general parameters for the Elbro term are biased towards the hydrocarbon and halocarbon data, diverging from those regressed for other combinatorial terms using a dataset that is more evenly distributed across various types of chemicals. Whereas all sets of general and atom-specific dispersion parameters in Table \ref{tab:param_deviation} are implemented in all of the considered models, Table \ref{tab:param_deviation_cross} extends those with parameters for cross-interactions.
\begin{table}[htbp]
    \centering
    \caption{Parameter values and deviations for openCOSMO-RS dispersion parametrizations without cross-interaction parameters.}
    \begin{adjustbox}{width=\textwidth} 
    \begin{tabular}{lcccccc}
        \toprule
        \multirow{2}{*}{\shortstack[c]{Parameters and\\[1.5ex] Deviations (AAD)}} & \multicolumn{6}{c}{Parametrizations} \\
        \cmidrule(lr){2-7}
        & SG\_6 & SG\_7 & FH\_6 & FH\_7 & Elbro\_6 & Elbro\_7 \\
        \midrule
        \multirow{1}{*}{\(a_{eff}\) [Å\textsuperscript{2}]} & \multicolumn{2}{c}{6.115} & \multicolumn{2}{c}{5.034} & 2.745619 & 2.773896 \\
         \(r_{av}\) [Å] & \multicolumn{6}{c}{0.5}\\
        \multirow{1}{*}{\(\alpha_{mf}\) [J/(mol$\cdot$Å\textsuperscript{2})/e\textsuperscript{2}]} & \multicolumn{2}{c}{7.584E+06} & \multicolumn{2}{c}{7.592E+06} & 1.210E+07 & 1.198E+07 \\
        \(f_{corr}\) & \multicolumn{6}{c}{2.4}\\
        \multirow{1}{*}{\( c_{hb}\) [J/(mol$\cdot$Å\textsuperscript{2})/e\textsuperscript{2}]} & \multicolumn{2}{c}{3.093E+07} & \multicolumn{2}{c}{3.094E+07} & \multicolumn{2}{c}{3.093E+07} \\

        \(c_{hb}^T\) & \multicolumn{6}{c}{1.5}\\
        \multirow{1}{*}{\(\sigma_{hb}\) [e/Å\textsuperscript{2}]} & \multicolumn{2}{c}{0.007876} & \multicolumn{2}{c}{0.007276} & \multicolumn{2}{c}{0.007876} \\
        \multirow{1}{*}{Astd [Å\textsuperscript{2}]} & \multicolumn{2}{c}{41.89} & & & & \\
        $\tau^{\text{vdW}}_{\text{C(sp}^3)}$ [J\textsuperscript{0.5}/Å] & 11.193 & 9.425 & 14.123 & 10.577 & 17.675 & 18.221 \\
        $\tau^{\text{vdW}}_{\text{C(sp}^2)}$ [J\textsuperscript{0.5}/Å] & & 10.235 & & 11.480 & & 19.054 \\
        $\tau^{\text{vdW}}_{\text{H}}$ [J\textsuperscript{0.5}/Å] & 10.041 & 9.021 & 12.581 & 10.300 & 15.636 & 17.506 \\
        $\tau^{\text{vdW}}_{\text{F}}$ [J\textsuperscript{0.5}/Å] & 3.240 & 1.977 & 5.319 & 2.522 & 8.914 & 10.260 \\
        $\tau^{\text{vdW}}_{\text{Cl}}$ [J\textsuperscript{0.5}/Å] & 11.865 & 10.647 & 14.575 & 11.735 & 18.289 & 19.669 \\
        $\tau^{\text{vdW}}_{\text{Br}}$ [J\textsuperscript{0.5}/Å] & 17.602 & 16.414 & 20.796 & 17.545 & 25.965 & 27.218 \\
        $\tau^{\text{vdW}}_{\text{I}}$ [J\textsuperscript{0.5}/Å] & 19.578 & 18.236 & 22.823 & 20.031 & 26.716 & 28.244 \\
        \midrule
        AAD\textsubscript{TOT} & 0.0473 & 0.0469 & 0.0466 & 0.0460 & 0.0491 & 0.0489 \\
        AAD\textsubscript{IDAC} & 0.1949 & 0.1929 & 0.2528 & 0.2521 & 0.1007 & 0.0998 \\
        AAD\textsubscript{LLE} & 0.1344 & 0.1328 & 0.0964 & 0.0937 & 0.1055 & 0.1045 \\
        AAD\textsubscript{VLE} & 0.0296 & 0.0294 & 0.0327 & 0.0324 & 0.0390 & 0.0389 \\
        \bottomrule
    \end{tabular}
    \end{adjustbox}
    \label{tab:param_deviation}

\end{table}

\begin{figure}
    \centering
    \includegraphics[width=0.75\linewidth]{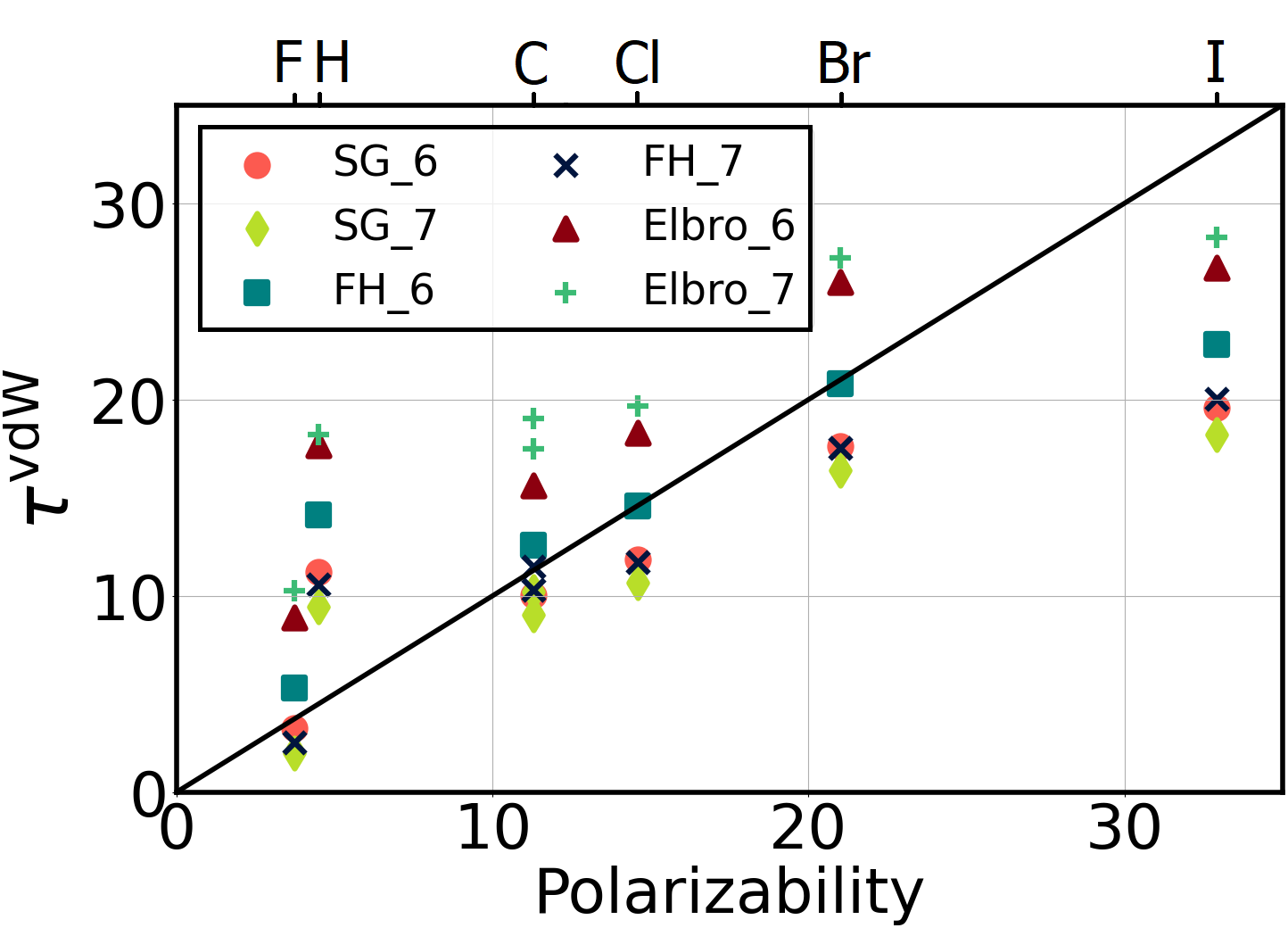}
    \caption{Comparison between atomic polarizabilities from \citet{Schwerdtfeger20192018Table} and atom-specific dispersion parameters (\(\tau^{\text{vdW}}\)).}
    \label{fig:polarizability_tau}
\end{figure}

\begin{table}[htbp]
    \caption{Cross-interaction parameter values and deviations for openCOSMO-RS dispersion parametrizations with cross-interaction parameters.}
    \centering
    \begin{adjustbox}{width=\textwidth} 
    \begin{tabular}{lcccccc}
        \toprule
        \multirow{2}{*}{\shortstack[c]{Parameters\\[1ex]and Deviations (AAD)}} & \multicolumn{6}{c}{Parametrizations} \\
        \cmidrule(lr){2-7}
        & SG\_6\_cross & SG\_7\_cross & FH\_6\_cross & FH\_7\_cross & Elbro\_6\_cross & Elbro\_7\_cross \\
        \midrule
        $\text{k}_{\text{H-C(sp}^3)}$ & 0.18159 & 0.01833 & 0.24871 & 0.00159 & 0.15927 & -0.81875 \\
        $\text{k}_{\text{H-C(sp}^2)}$ & & -0.11976 & & 0.15274 & & -0.34582 \\
        $\text{k}_{\text{H-F}}$ & 0.06897 & 0.66191 & 0.09735 & 0.74790 & 0.07372 & -0.17848 \\
        $\text{k}_{\text{H-Cl}}$ & 0.22346 & 0.25723 & 0.16732 & 0.14724 & 0.14310 & -0.04194 \\
        $\text{k}_{\text{H-Br}}$ & -0.10650 & -0.24431 & 0.16923 & 0.16498 & 0.10780 & -0.25552 \\
        $\text{k}_{\text{H-I}}$ & 0.56261 & 0.29858 & 0.29177 & 0.22584 & 0.36641 & 0.11027 \\
        $\text{k}_{\text{C(sp}^3)-\text{C(sp}^2)}$ & & 0.13688 & & 0.01047 & & -0.07518 \\
        $\text{k}_{\text{C(sp}^3)-\text{F}}$ & 0.25838 & -0.73798 & 0.24347 & -0.94343 & 0.07947 & -0.63680 \\
        $\text{k}_{\text{C(sp}^2)-\text{F}}$ & & -0.38252 & & 0.01254 & & -0.14392 \\
        $\text{k}_{\text{C(sp}^3)-\text{Cl}}$ & -0.09500 & -0.19281 & -0.01799 & -0.10530 & -0.03320 & -0.33904 \\
        $\text{k}_{\text{C(sp}^2)-\text{Cl}}$ & & -0.19363 & & -0.03107 & & -0.11235 \\
        $\text{k}_{\text{C(sp}^3)-\text{Br}}$ & -0.13948 & 0.07020 & -0.08836 & -0.20372 & -0.09358 & -0.04681 \\
        $\text{k}_{\text{C(sp}^2)-\text{Br}}$ & & -0.15348 & & 0.00222 & & -0.04118 \\
        $\text{k}_{\text{C(sp}^3)-\text{I}}$ & -0.47451 & -0.25369 & -0.25303 & -0.21984 & -0.30310 & -0.45880 \\
        $\text{k}_{\text{C(sp}^2)-\text{I}}$ & & -0.42763 & & -0.37158 & & -0.28746 \\
        $\text{k}_{\text{F-Cl}}$ & -0.02230 & -0.14832 & -0.03033 & -0.10208 & -0.01986 & -0.02789 \\
        $\text{k}_{\text{F-Br}}$ & -0.40404 & -0.33715 & -0.01564 & -0.31896 & -0.11410 & 0.02610 \\
        $\text{k}_{\text{F-I}}$ & -0.12331 & 0.02270 & -0.19483 & 0.05881 & -0.12197 & -0.11127 \\
        $\text{k}_{\text{Cl-Br}}$ & 0.11674 & 0.00928 & -0.00186 & 0.06092 & 0.04101 & -0.02772 \\
        $\text{k}_{\text{Cl-I}}$ & -0.03422 & -0.03802 & -0.01921 & -0.02616 & -0.03001 & -0.02558 \\
        $\text{k}_{\text{Br-I}}$ & 0.08028 & -0.04237 & 0.01017 & 0.08432 & 0.05495 & 0.02172 \\
        \midrule
        $\text{AAD}_\text{TOT}$ & 0.0431 & 0.0424 & 0.04015 & 0.03878 & 0.04004 & 0.03822 \\
        $\text{AAD}_\text{IDAC}$ & 0.1838 & 0.1898 & 0.22506 & 0.22706 & 0.07917 & 0.08143 \\
        $\text{AAD}_\text{LLE}$ & 0.1326 & 0.1268 & 0.09632 & 0.08888 & 0.10071 & 0.09645 \\
        $\text{AAD}_\text{VLE}$ & 0.0252 & 0.0250 & 0.02595 & 0.02539 & 0.02962 & 0.02806 \\
        \bottomrule
    \end{tabular}
    \end{adjustbox}
    \label{tab:param_deviation_cross}
\end{table}
For regressions, we utilized the differential evolution algorithm provided in the SciPy Python package \citep{2020SciPy-NMeth}. While this algorithm does not require initial estimates and avoids using gradient methods, it explores a large sample parameter space, resulting in longer optimization times. However, its strength lies in its ability to identify the global minimum within the parameter space more effectively than conventional gradient methods. Furthermore, the minimizations can be constrained using bounds and dependencies between parameters. Therefore, to impose physically meaningful constraints, we used the values of atomic polarizabilities listed in Table \ref{tab:polarizabilities}. Consequently, the atom-specific parameters align well with these polarizability values, as demonstrated in Figure \ref{fig:polarizability_tau}, with parametrizations involving the FH combinatorial term being the closest match. 
\begin{table}[htbp]
    \centering
    \caption{Atomic polarizabilities from \citet{Schwerdtfeger20192018Table}.}
    \begin{tabular}{cc}
    \hline
    Atomic symbol & Polarizability [a.u.\(^3\) ] \\
    \hline
    H & 4.5 \\
    C & 11.3 \\
    F & 3.74 \\
    Cl & 14.6 \\
    Br & 21 \\
    I & 32.9 \\
    \hline
    \end{tabular}
    \label{tab:polarizabilities}
\end{table}

\begin{figure}
    \centering
    \includegraphics[width=0.45\linewidth]{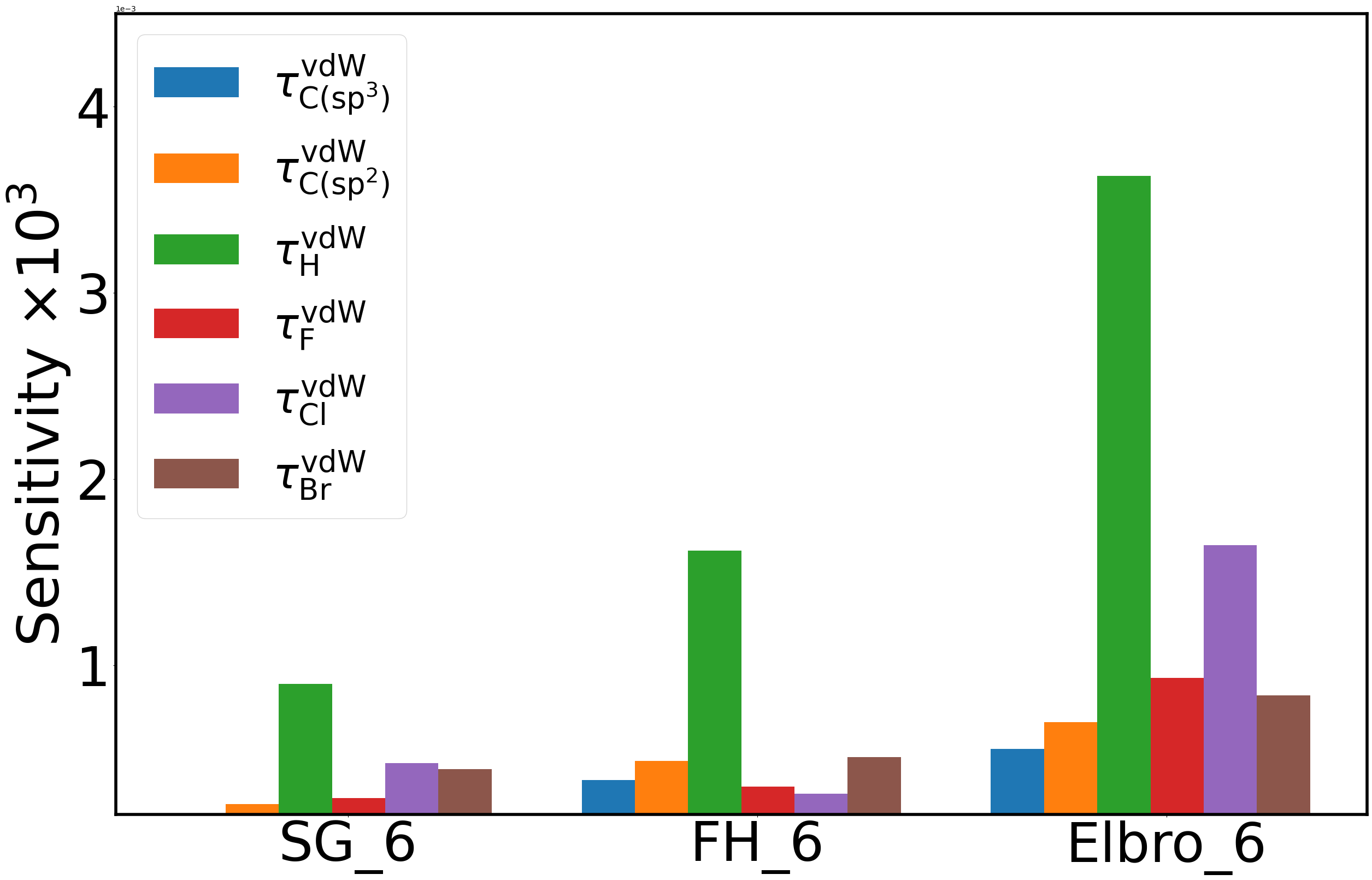}
    \includegraphics[width=0.45\linewidth]{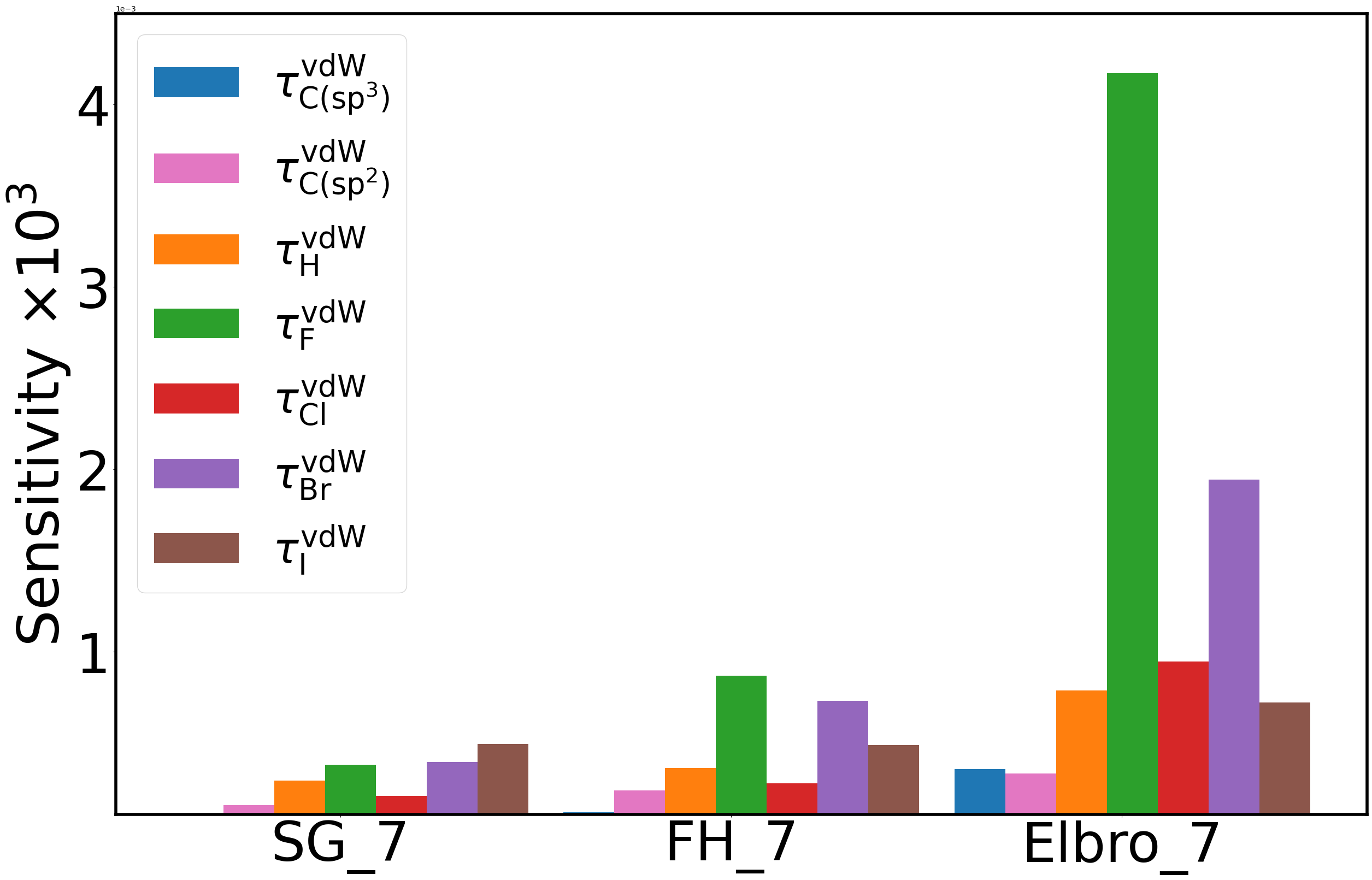}
    \caption{Local sensitivities for atom-specific parameters estimated for all types of phase equilibrium data.}
    \label{fig:sensitivity_TOT_no_cross}
\end{figure}

Additionally, we estimated the local sensitivities \(S_x\) of the \(\tau^\text{vdW}
\) parameters using a backward difference method:
\begin{equation}
    S_x = \frac{1}{N_P}\frac{OF(x-\epsilon)-OF(x)}{\epsilon},
\label{eq:backward_sensitivity}
\end{equation}
where \textit{x} is the evaluated parameter, with the others being fixed, \textit{OF} is the value of the objective function at \textit{x}, and \(\epsilon\) is the increment, which was set to 5\% of the parameter value. Due to a scarcity of data with Br and I atoms compared to the data with F and Cl atoms, we weighted the sensitivity with the number of data points \(N_P\) across the data containing these atoms.  
An example of such sensitivity analysis is demonstrated in Figure \ref{fig:sensitivity_TOT_no_cross} for models without cross-atom parameters. Notably, the highest sensitivities are mostly attributed to the F-atom parameter, while H and C atoms parameters show low sensitivities. Interestingly, all sensitivities for Elbro\_6 and Elbro\_7 models are much higher than those estimated with other combinatorial terms. 

To assess the performance of the modified models, Tables \ref{tab:param_deviation} and \ref{tab:param_deviation_cross} also provide the Average Absolute Deviations (AAD) estimated using the following equations:
\begin{equation}
\mathrm{AAD_{LLE}} = \frac{1}{N_p}\sum^{N_s}\sum_{i}^{Nc}\frac{\left|\ln(K_i^{\text{calc}}) - \ln(K_i^{\text{exp}}) \right|}{N_c},
\label{eq:AAD_LLE}
\end{equation}
\begin{equation}
\mathrm{AAD_{VLE}} = \frac{1}{N_p}\sum^{N_s}\sum_{i}^{Nc}\frac{\left|\ln(\gamma_i^{calc}) - \ln(\gamma_i^{exp}) 
\right|}{N_c},
\label{eq:AAD_VLE}
\end{equation}
\begin{equation}
\mathrm{AAD_{IDAC}} = \frac{1}{N_p}\sum^{N_s}\left|\ln(\gamma_\infty^{calc}) - \ln(\gamma_\infty^{exp}) 
\right|,
\label{eq:AAD_IDAC}
\end{equation}
and \(\text{AAD}_\text{TOT}\) calculated as the sum of all the absolute deviations divided by the total number of the data points across all types of data. 
\subsection{Evaluation of combinatorial terms for IDAC predictions}
\label{sec:combinatorial_term_discussion}
\begin{figure}
    \centering
    \includegraphics[width=0.45\linewidth]{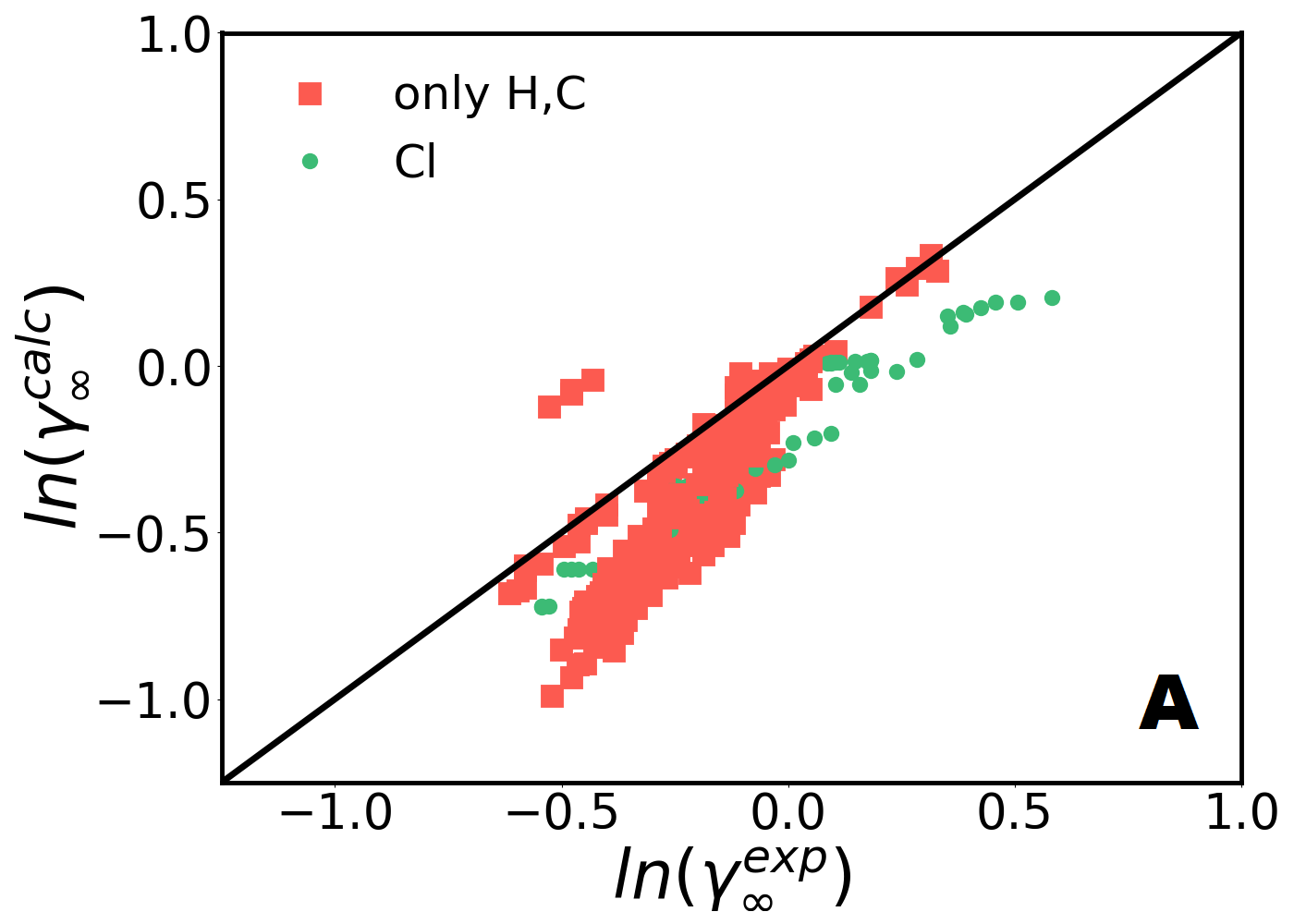}
    \includegraphics[width=0.45\linewidth]{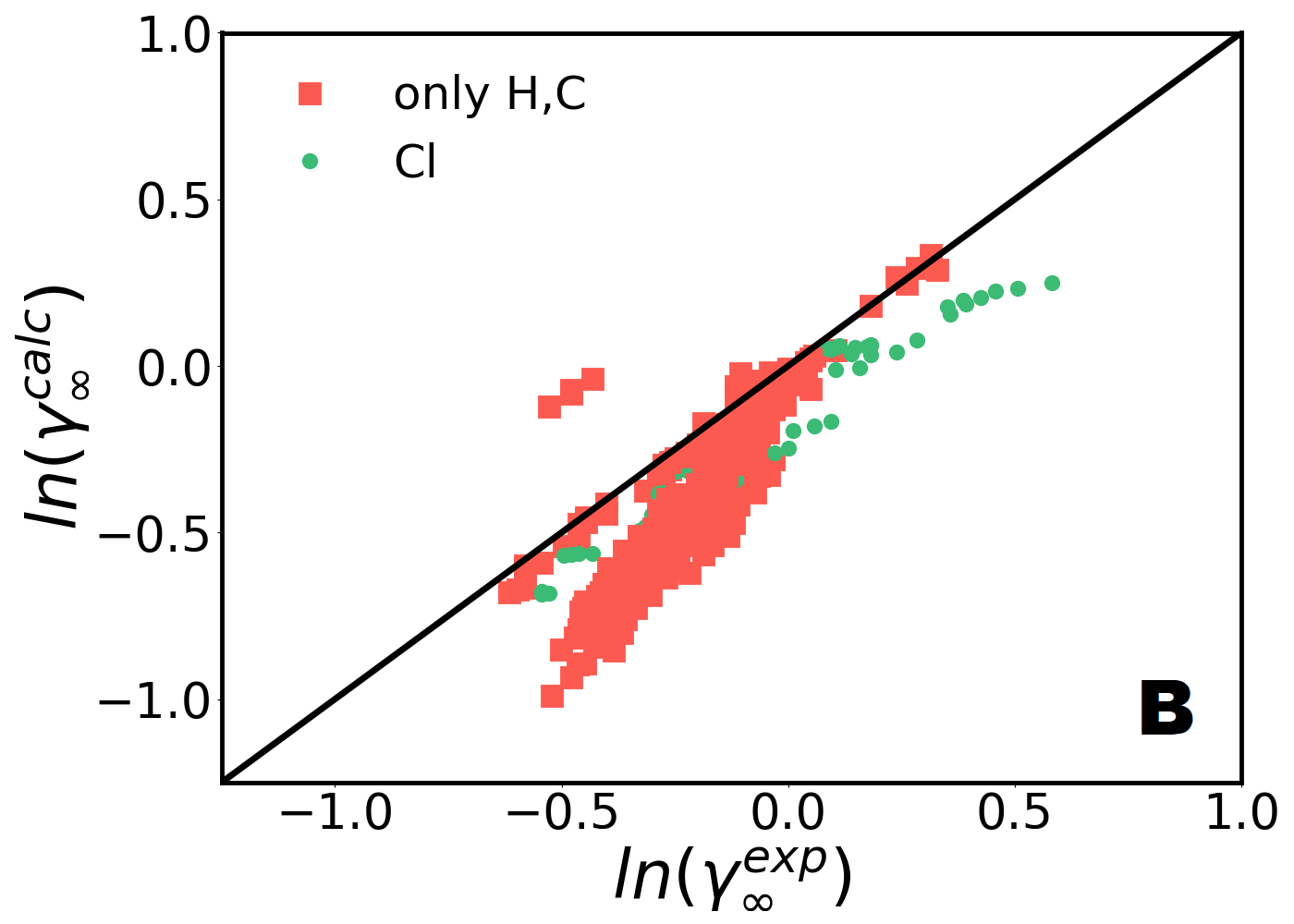}
    \caption{Parity plots comparing predicted values (calc) with experimental data (exp) for IDACs without the dispersion term (\textbf{A}) or using SG\_6 parametrization (\textbf{B}). The symbols are colored according to the type of atoms in the data.}
    \label{fig:IDAC_parity}
\end{figure}

\begin{figure}
    \centering
    \includegraphics[width=0.3\linewidth]{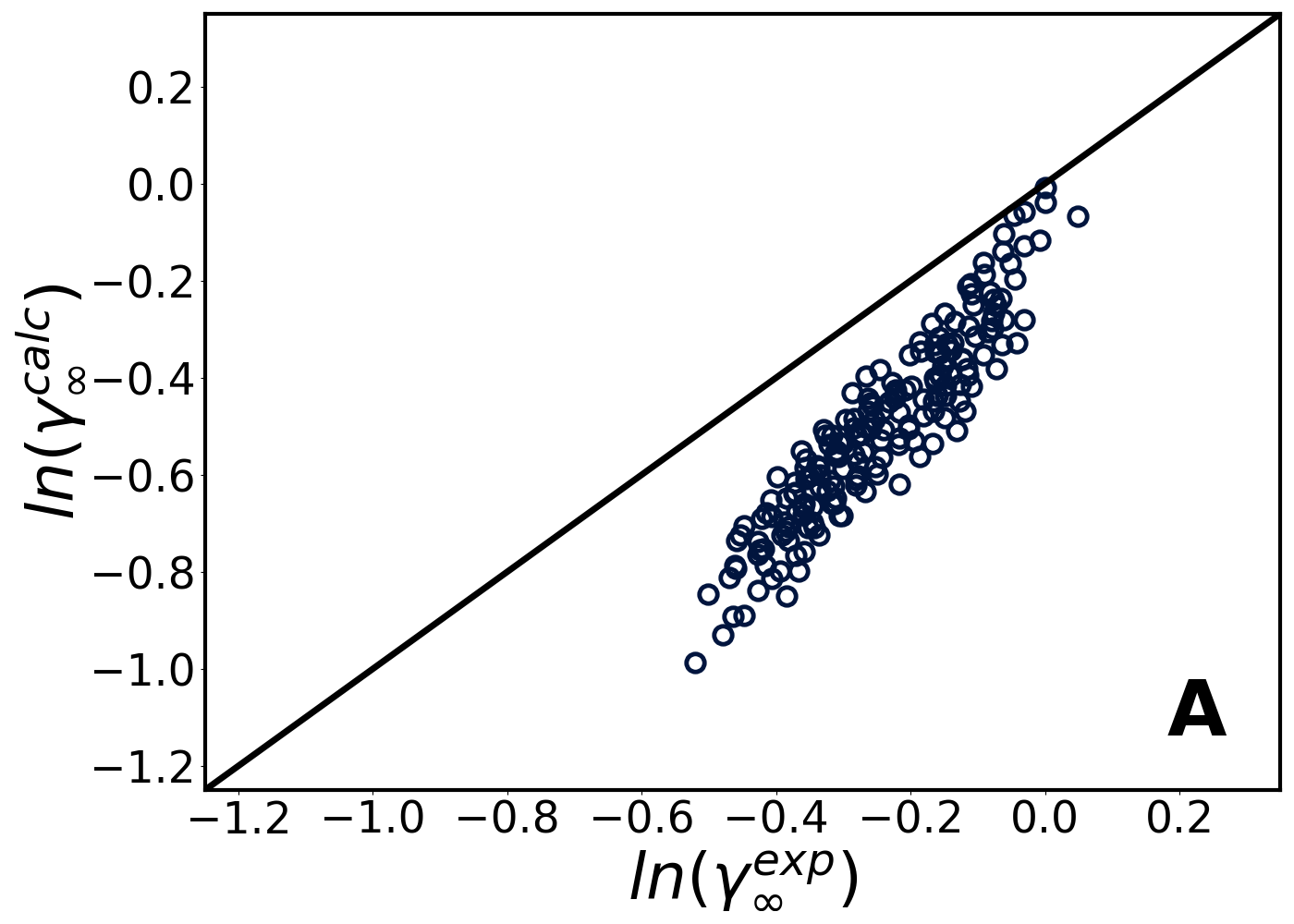}
    \includegraphics[width=0.3\linewidth]{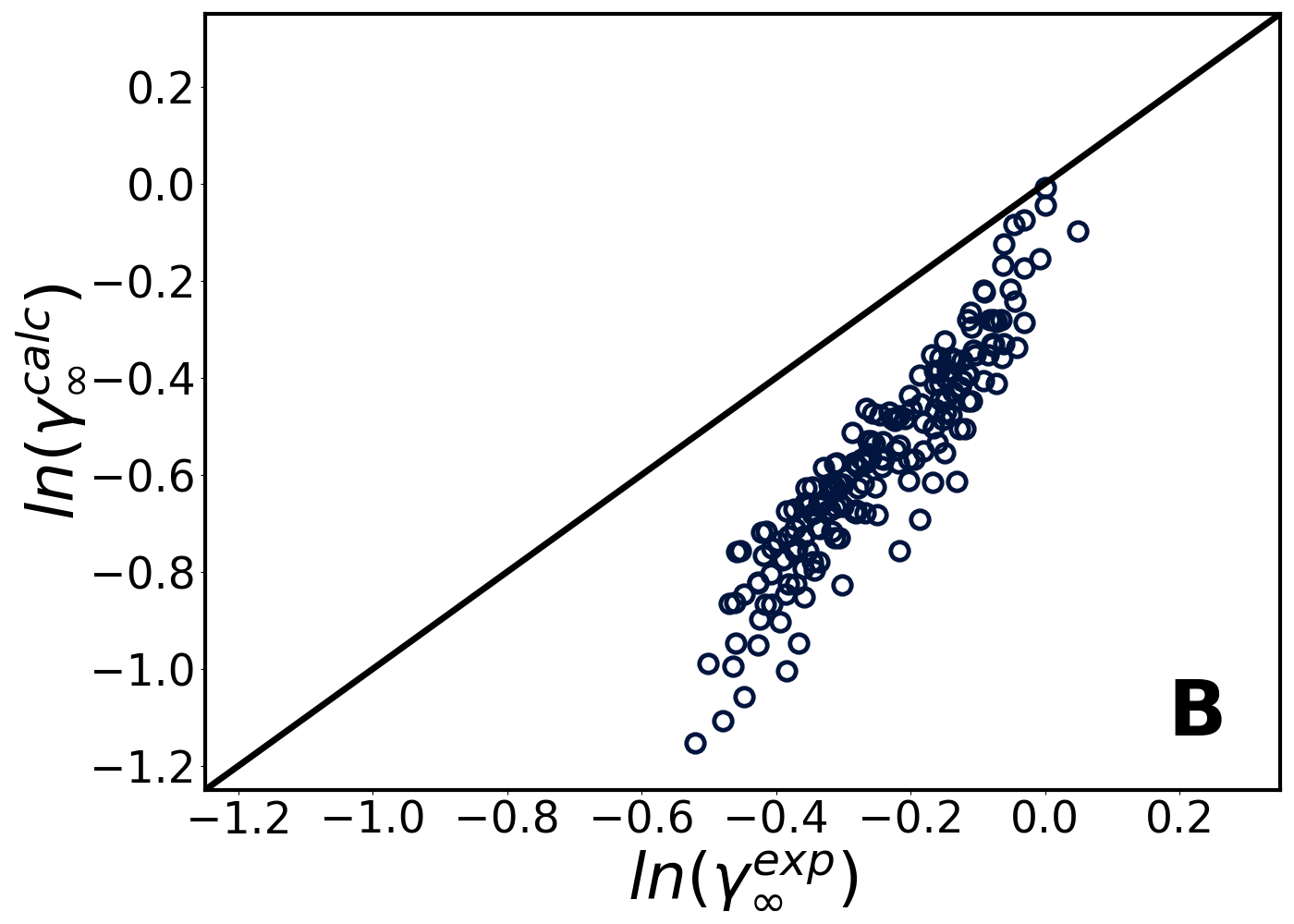}
    \includegraphics[width=0.3\linewidth]{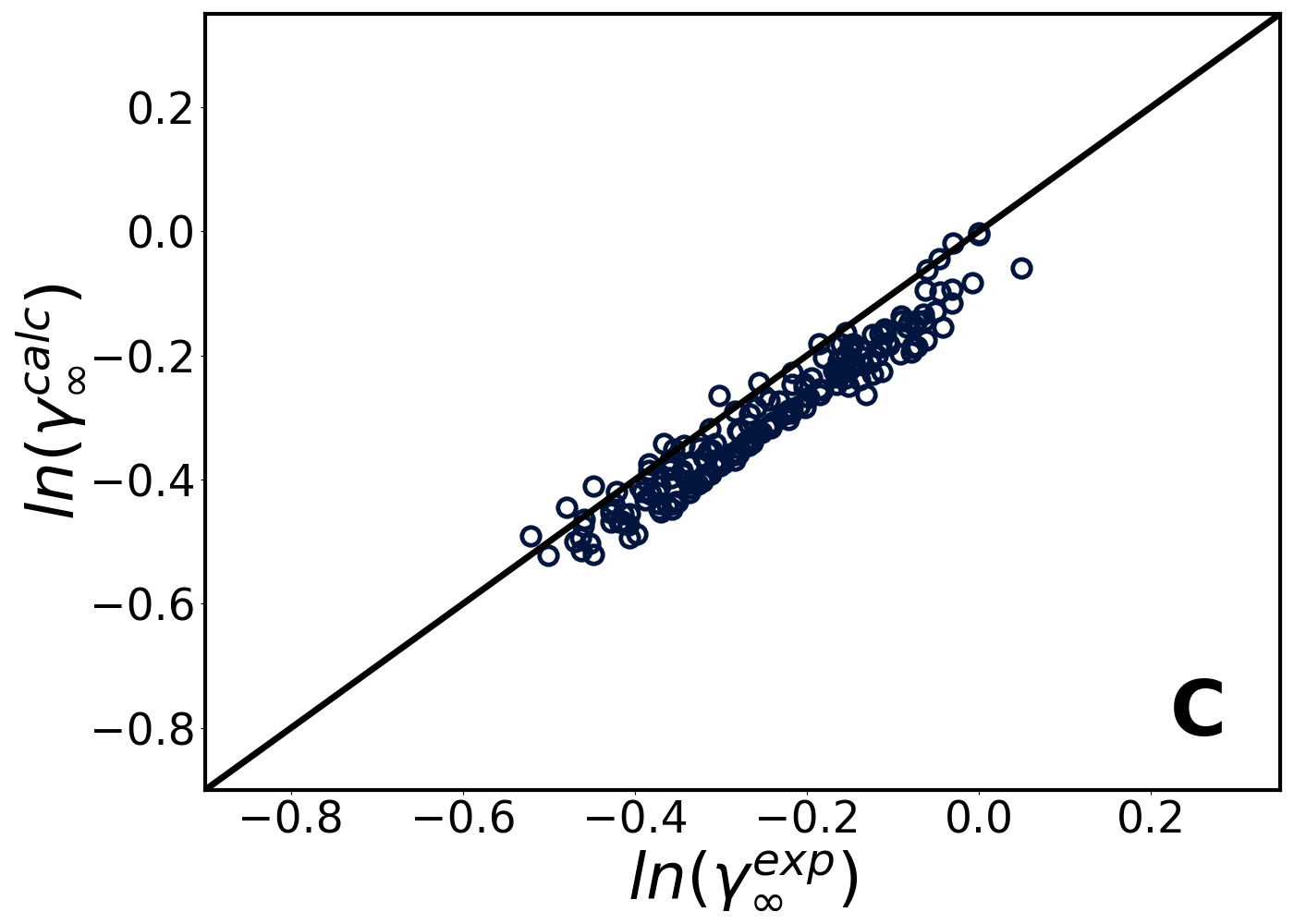}
    \caption{Parity plots for predicted IDAC (calc) versus experimental data (exp) using one of the following combinatorial terms solely: SG (\textbf{A}), FH (\textbf{B}) or Elbro (\textbf{C}). }
    \label{fig:IDACs_combinatorial}
\end{figure}

\begin{figure}
    \centering
    \includegraphics[width=0.45\linewidth]{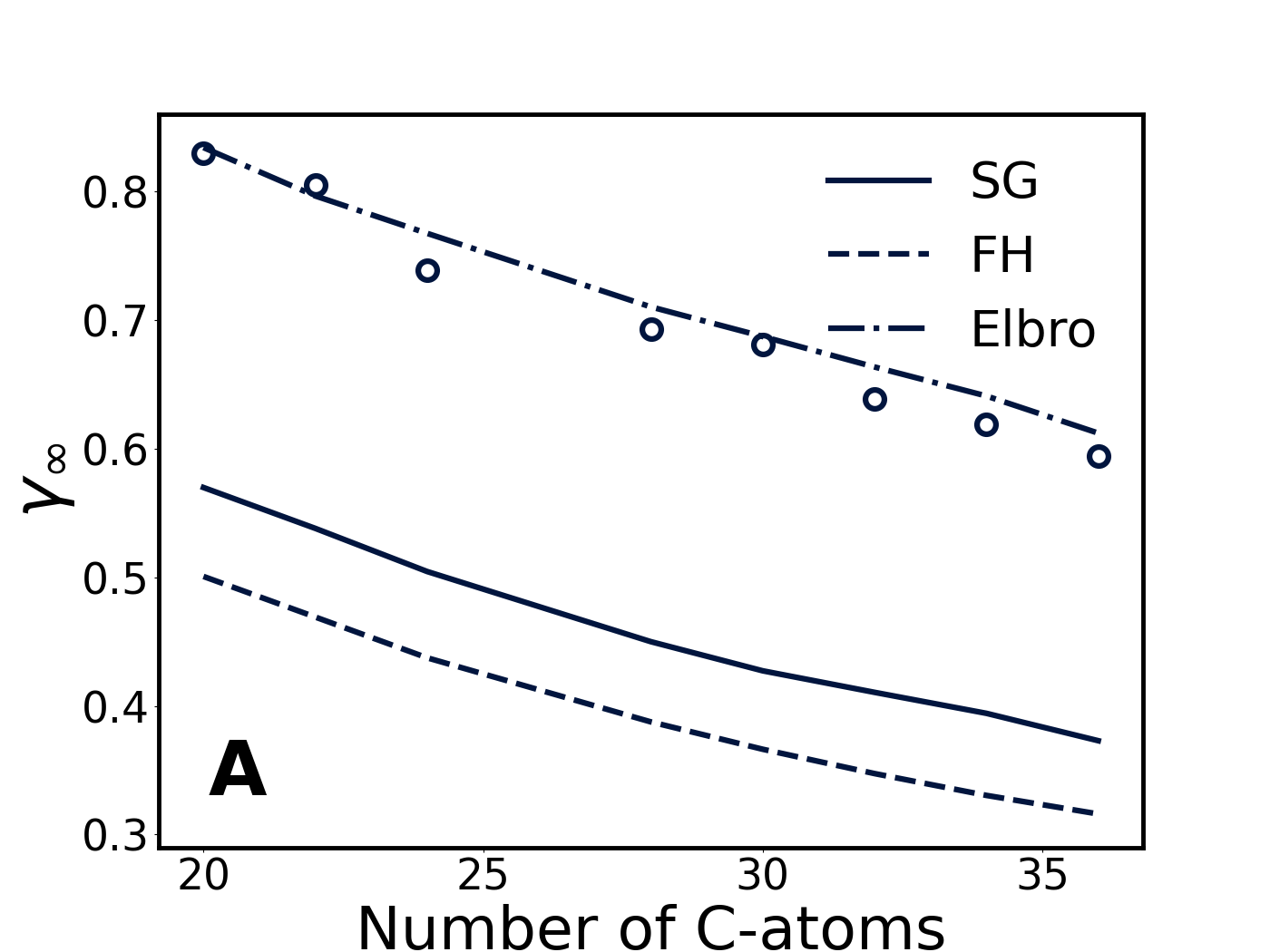}
    \includegraphics[width=0.45\linewidth]{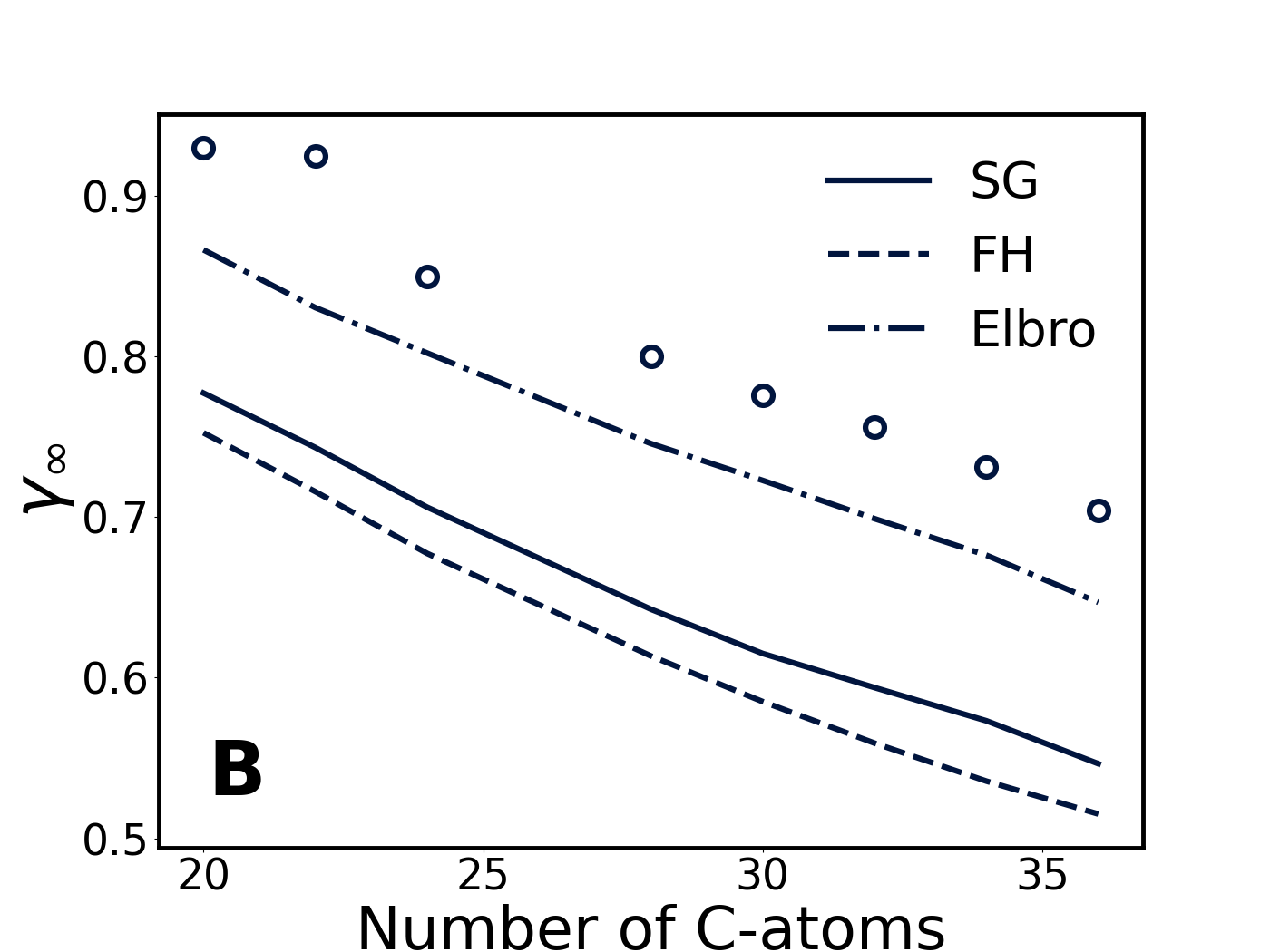}
    \caption{The experimental (\textbf{o}) and predicted by SG, FH and Elbro terms IDAC values of n-butane (\textbf{A}) and 4-methylheptane (\textbf{B}) diluted in long-chain alkanes with carbon number indicated of the x-axis.}
    \label{fig:IDACs_combinatorial_carbon_number}
\end{figure}

For the IDAC data, the dispersive contribution does not significantly improve openCOSMO-RS predictions as shown in Figure \ref{fig:IDAC_parity}, since most of the collected data consists of asymmetric hydrocarbon mixtures with only a few halocarbon datasets. Typically, both models tend to underestimate the activity coefficients. There are, however, a few outliers to this trend, specifically in the benzene/toluene - hexadecane systems. As shown in Figure S1, the IDACs of benzene and toluene sharply decrease with increasing temperature, forming an S-shaped curve. This peculiar trend may raise questions about the quality of the data.
For highly asymmetric alkane mixtures, the combinatorial term plays a crucial role. Thus, we evaluated three combinatorial terms: SG, FH and Elbro on the IDAC data of alkane mixtures collected by \citet{Soares2011TheModelsb}. It should be noted that the evaluated IDAC data consists of short-chain solutes in long-chain solvents only. As shown in Figures \ref{fig:IDACs_combinatorial} and \ref{fig:IDACs_combinatorial_carbon_number}, the Elbro term outperforms both SG and FH, which show comparable performance, with FH being the most inferior. Remarkably, the evidence from molecular dynamic studies \citep{Iwai2010TestModels} and polymers research \citep{Silva2023TheBehavior} indicates that the Elbro term is in better agreement with Monte Carlo
simulations and leads to more accurate predictions for polymer solutions, when coupled with COSMO-RS implemented in COSMOtherm. The performance of the FH and Elbro terms and their modifications for asymmetric alkanes and polymer mixtures has been previously investigated by \citet{Kontogeorgis1994ImprovedModels,Kontogeorgis1997ImprovedFunction, Kouskoumvekaki2002AnSolutions}.  It was found that the Elbro term performs well for short-chain solutes in long-chain solvents but not for long-chain solutes in short-chain solvents. \citet{Kouskoumvekaki2002AnSolutions} evaluated the performance of the Elbro term and suggested that the inaccessible hard-core volume might be higher than the van der Waals volume currently used. They increased its value by a factor of 1.2, which led to improved results. Interestingly, the volumes of COSMO cavities are related to the van der Waals volumes by the same factor \citep{Klamt1998RefinementCOSMO-RS}. Since these COSMO cavity volumes were used in our calculations, this could account for the superior performance of the Elbro term for the evaluated set. As stated by \citet{Kontogeorgis1994ImprovedModels}, it is challenging to predict both the activity coefficients of short-chain solutes in long-chain solvents and long-chain solutes in short-chain solvents with a single model. Generally speaking, to select the optimal combinatorial term for openCOSMO-RS, a broader experimental dataset should be considered in the evaluation, which we plan to address in future research. It is also important to emphasize that the Elbro term comes with practical limitations, one of which is the necessity of additional experimental data on the density of pure components to calculate the temperature-dependent molar volume. 

\subsection{Evaluation of the parametrization approaches for VLE and LLE predictions}
\label{sec:VLE_and_LLE_predictions}
\begin{figure}
    \centering
    \includegraphics[width=0.45\linewidth]{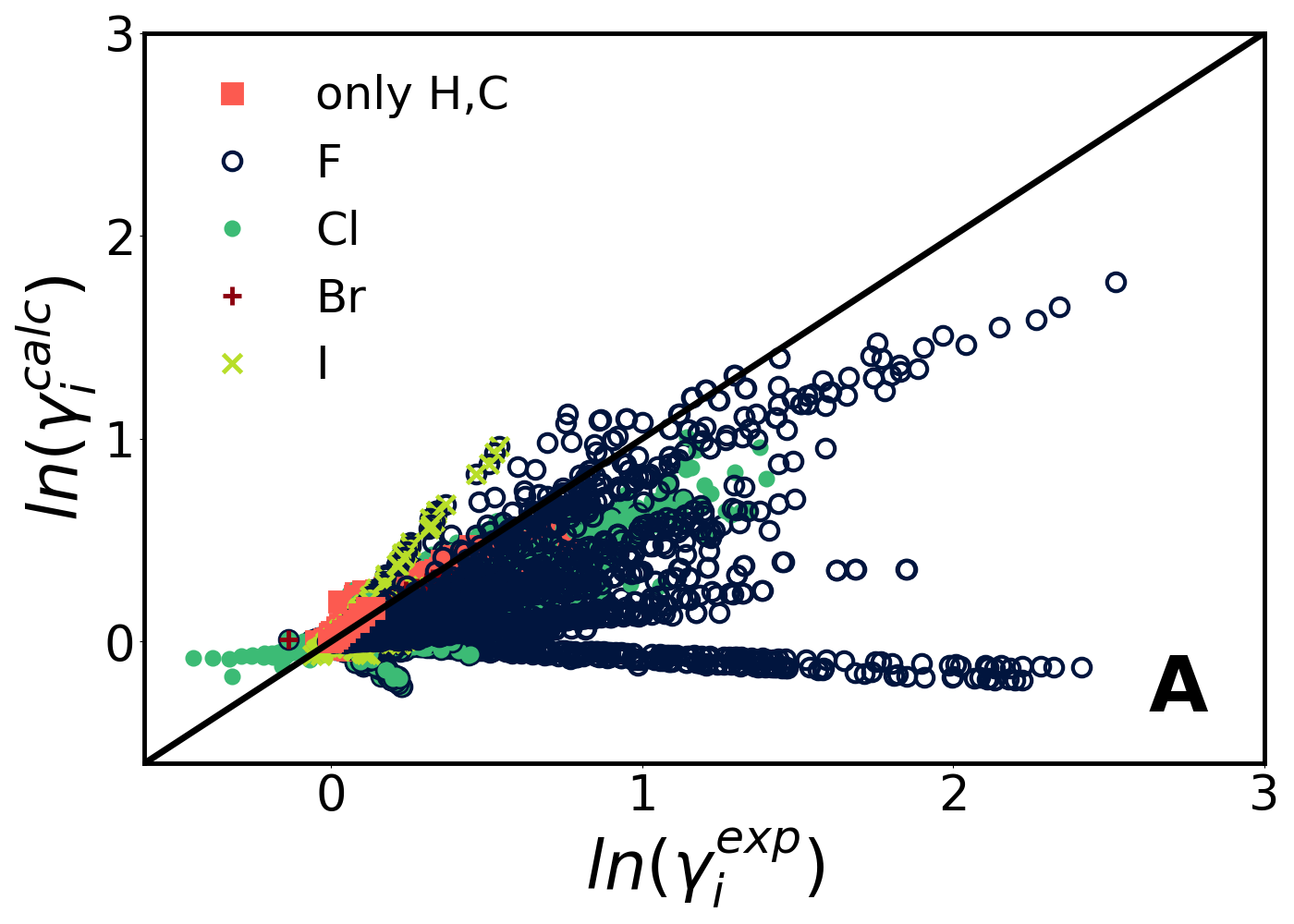}
    \includegraphics[width=0.45\linewidth]{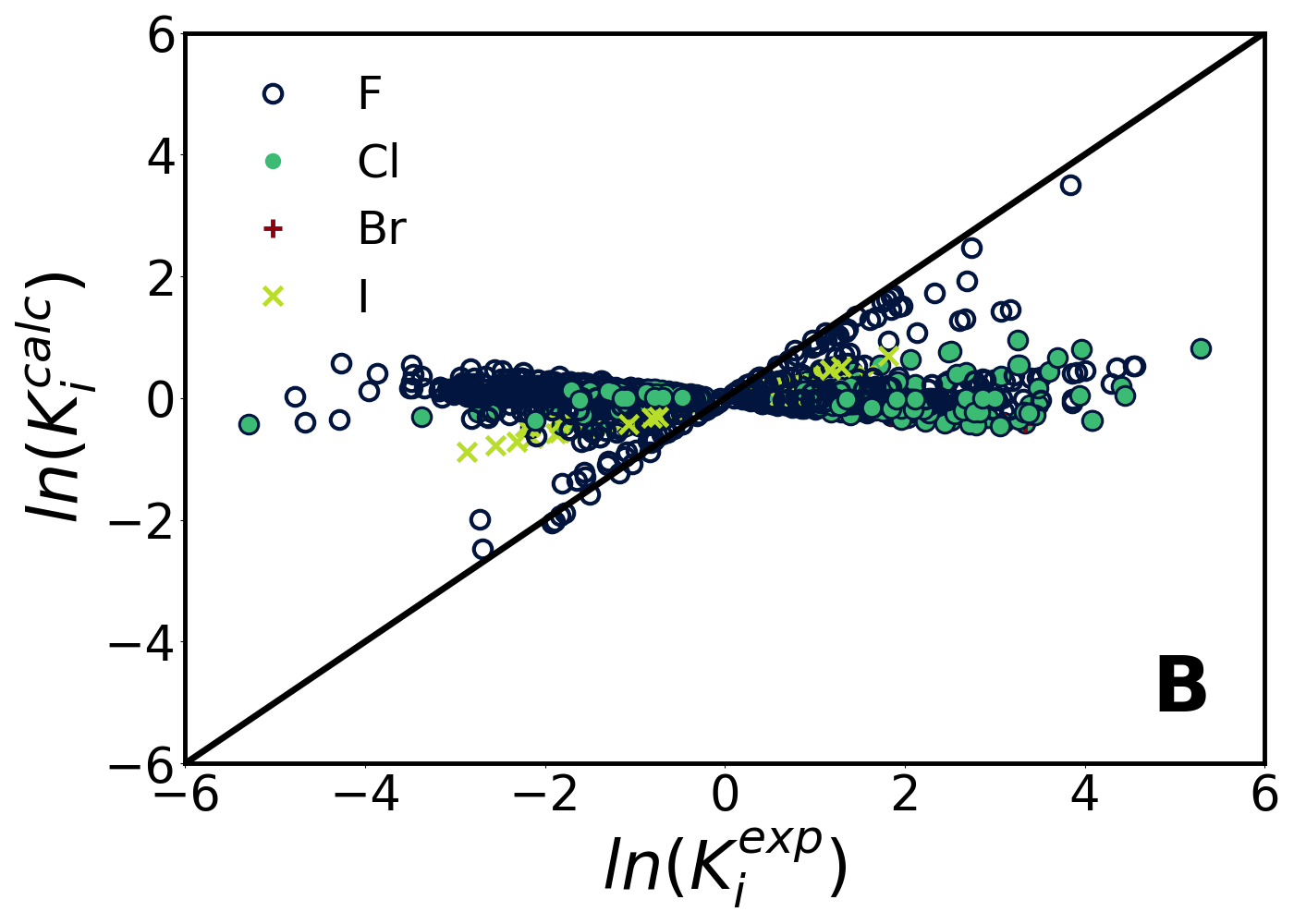}
    \includegraphics[width=0.45\linewidth]{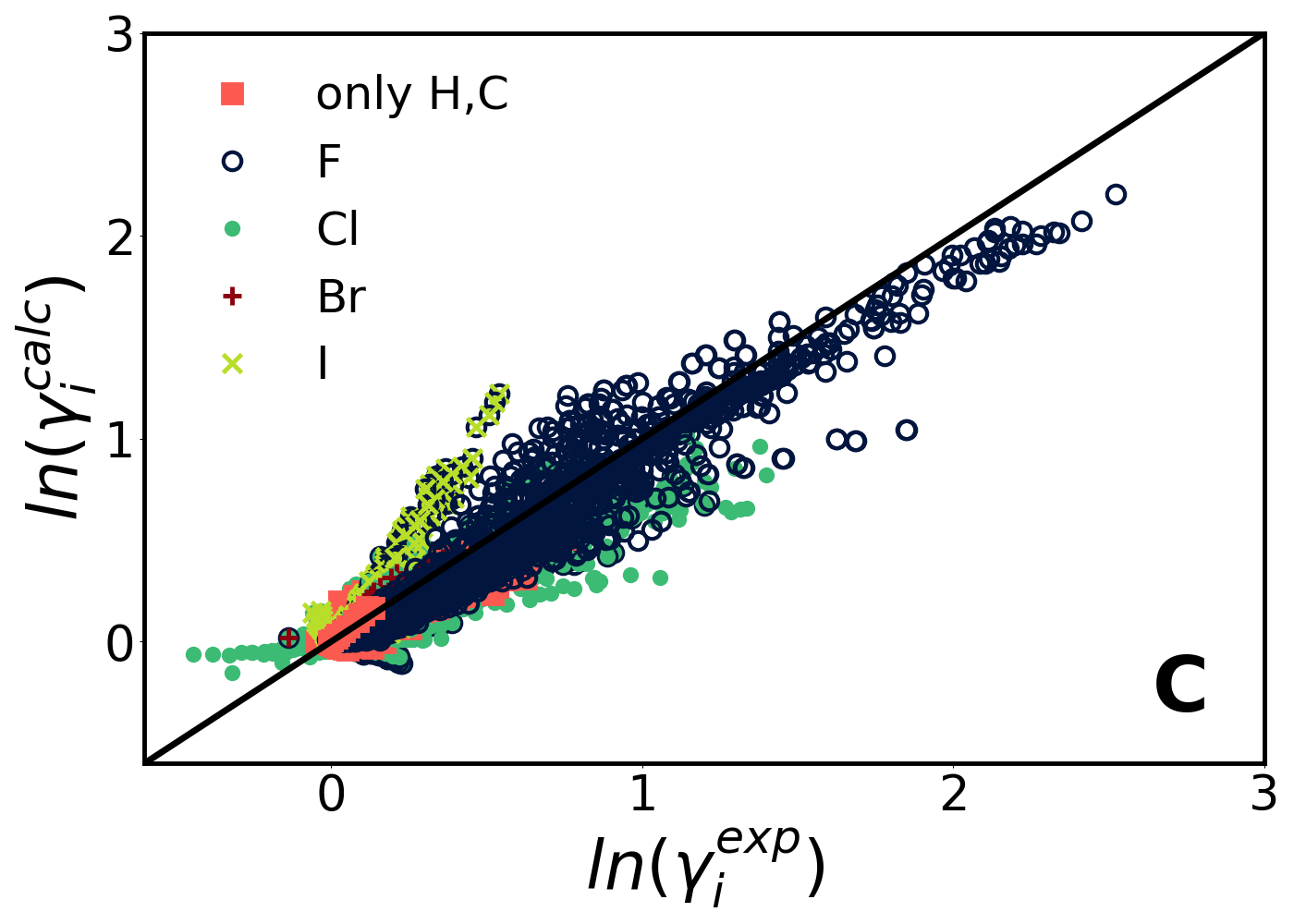}
    \includegraphics[width=0.45\linewidth]{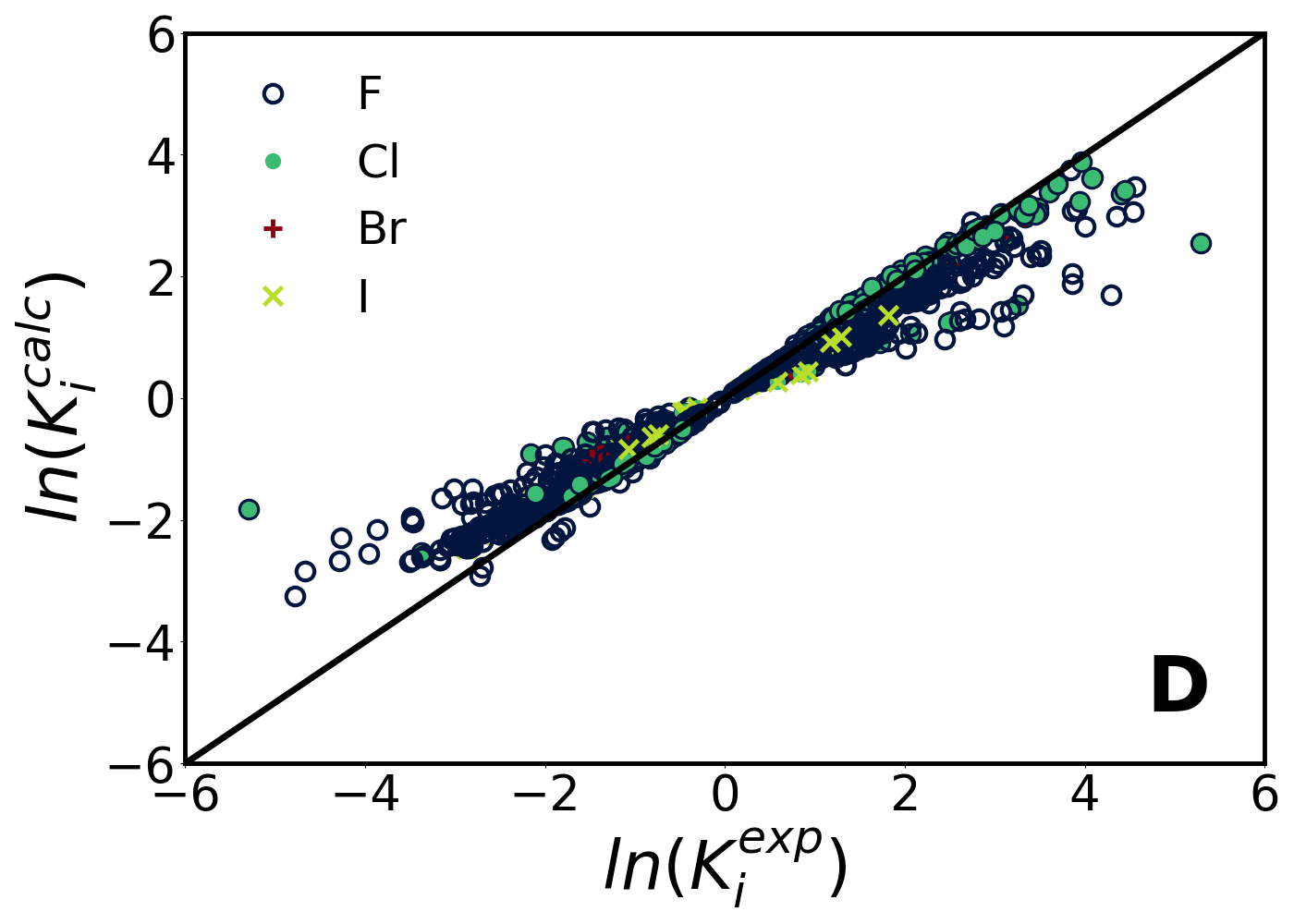} 
    \includegraphics[width=0.45\linewidth]{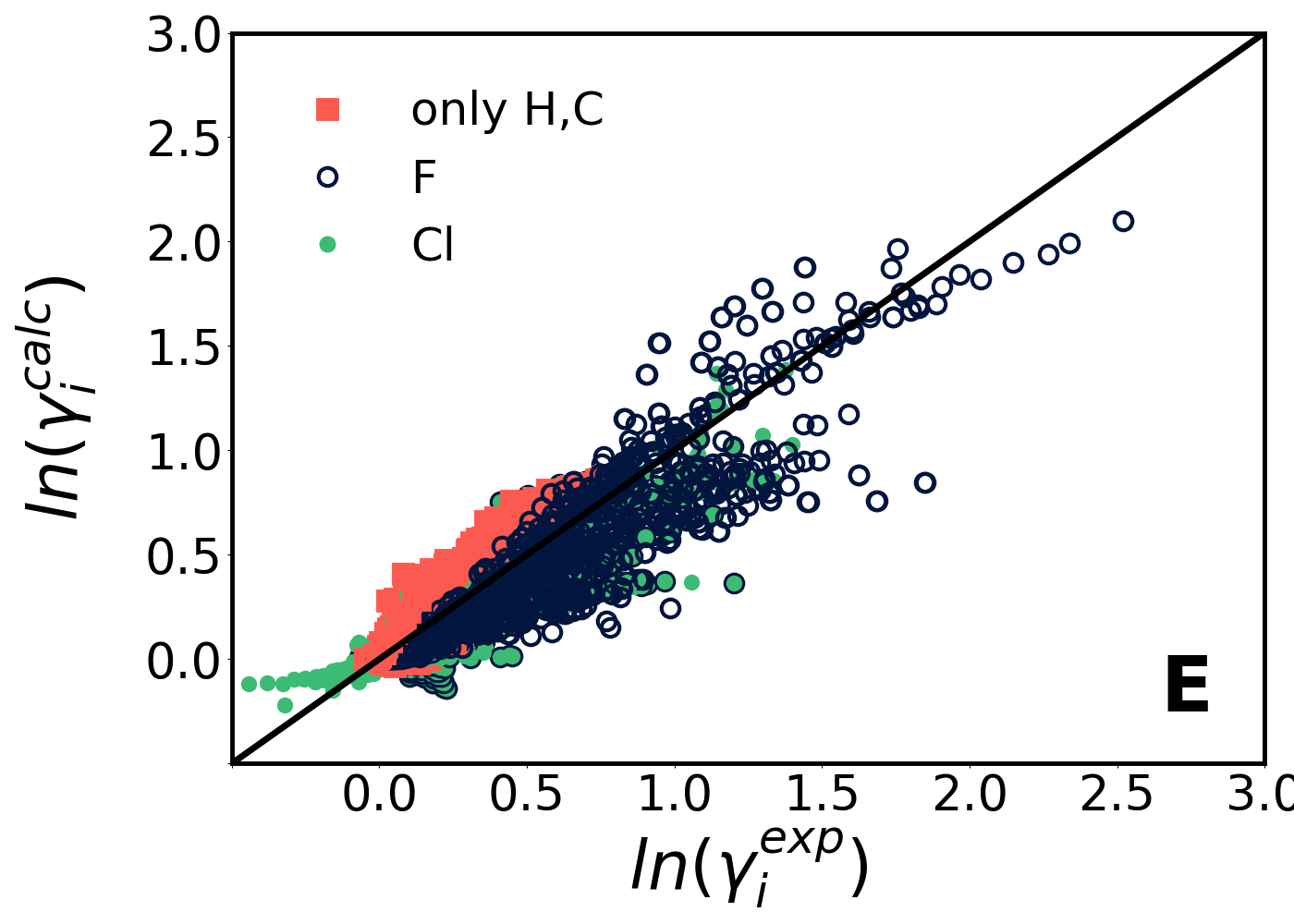} 
    \includegraphics[width=0.45\linewidth]{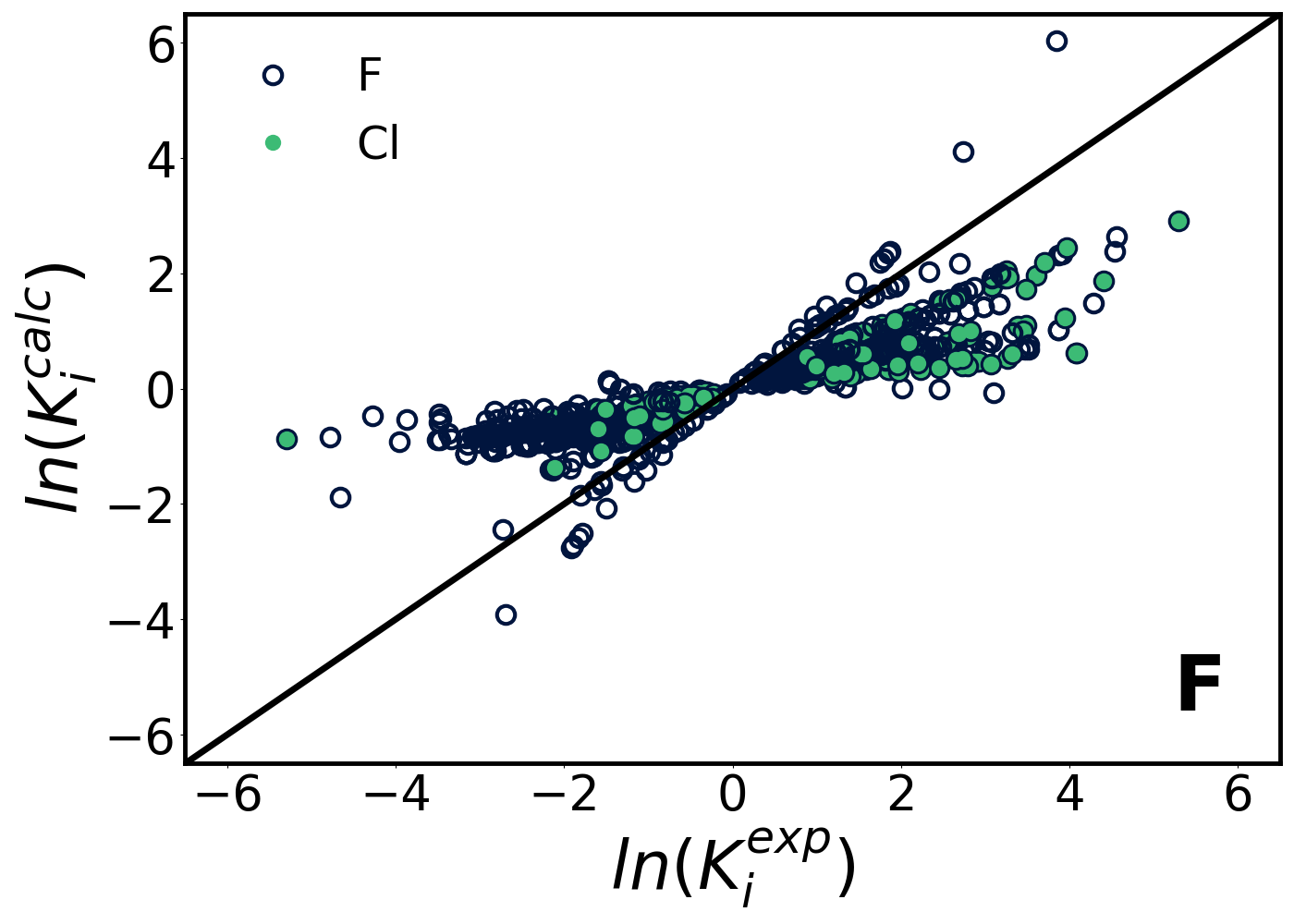} 
    \caption{Parity plots comparing predicted values (calc) with experimental data (exp) for VLE activity coefficients without the dispersion term (\textbf{A}), using SG\_6 parametrization (\textbf{C}) or COSMO-SAC-dsp (\textbf{E}), and for LLE partition coefficients without the dispersion term (\textbf{B}), using SG\_6 parametrization (\textbf{D}) or COSMO-SAC-dsp (\textbf{F}). The symbols are colored according to the type of atoms in the data.}
    \label{fig:VLE_LLE_parity}
\end{figure}

Disregarding the dispersion contribution leads to substantial deviations from the experimental data for both VLE and LLE, as one can observe in Figures \ref{fig:VLE_LLE_parity}A and \ref{fig:VLE_LLE_parity}B. Conversely, Figures \ref{fig:VLE_LLE_parity}C and \ref{fig:VLE_LLE_parity}D show that incorporating the dispersion term with a minimal number of adjustable parameters results in significant improvements in prediction accuracy. For comparison, the results obtained using the COSMO-SAC-dsp implementation by \citet{Bell2020ACOSMO-SAC} are presented in Figures \ref{fig:VLE_LLE_parity}E and \ref{fig:VLE_LLE_parity}F. Although, we used the same data for the evaluation, one can notice that there are no predictions for the brominated and iodinated compounds, as the corresponding dispersion parameters are not included in the model's parametrization. While COSMO-SAC-dsp demonstrates only slightly inferior but comparable performance for VLE data, it is notably inferior for LLE data. 

\begin{figure}
    \centering
    \includegraphics[width=0.5\linewidth]{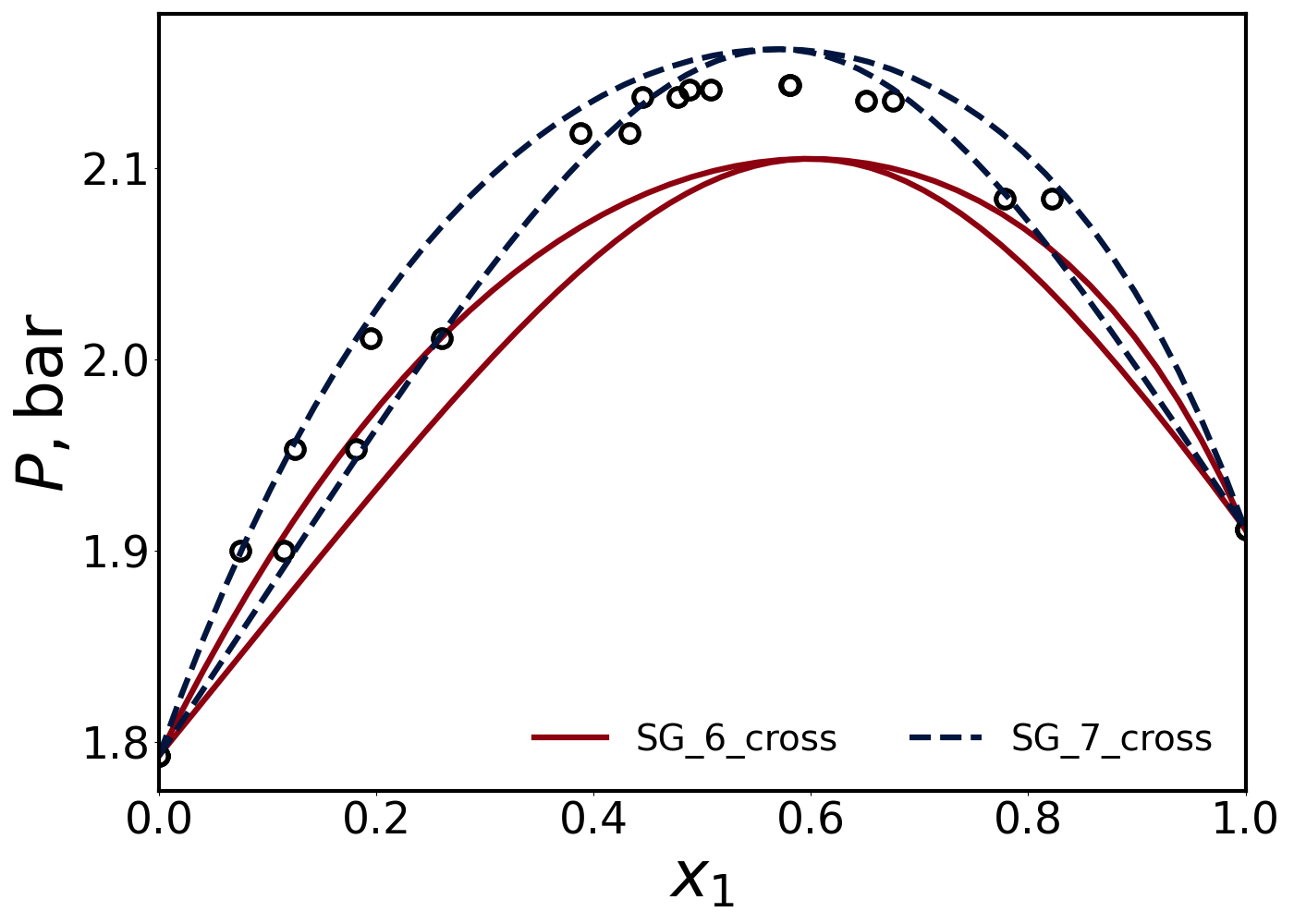}
    \caption{P-\textit{xy} phase diagram of trifluoroiodomethane - trans-1,3,3,3-tetrafluoropropene at 268.15 K. Experimental data \citep{Guo2012Vapour+liquid298.150K} is represented by symbols (\textbf{o}), while corresponding lines depict predictions made using various openCOSMO-RS dispersion parametrizations. These predictions illustrate the effect of differentiating between C-atoms with \(sp^3\) and \(sp^2\) hybridizations (SG\_7\_cross) compared to using a single dispersion parameter for C-atom (SG\_6\_cross).}
    \label{fig:new26_carbon_switch}
\end{figure}

The AAD values (Tables \ref{tab:param_deviation} and \ref{tab:param_deviation_cross}) for all the regressed models indicate that differentiating between \(sp^3\) and \(sp^2\) hybridizations of C-atoms does not significantly affect the predictions of phase equilibrium. While improvements can be seen for some systems, as shown in Figure \ref{fig:new26_carbon_switch}, overall, not much difference is observed on a large scale. Besides, we did not consider \(sp\) hybridization due to the absence of corresponding experimental data in our database. 

\begin{figure}
    \centering
    \includegraphics[width=0.45\linewidth]{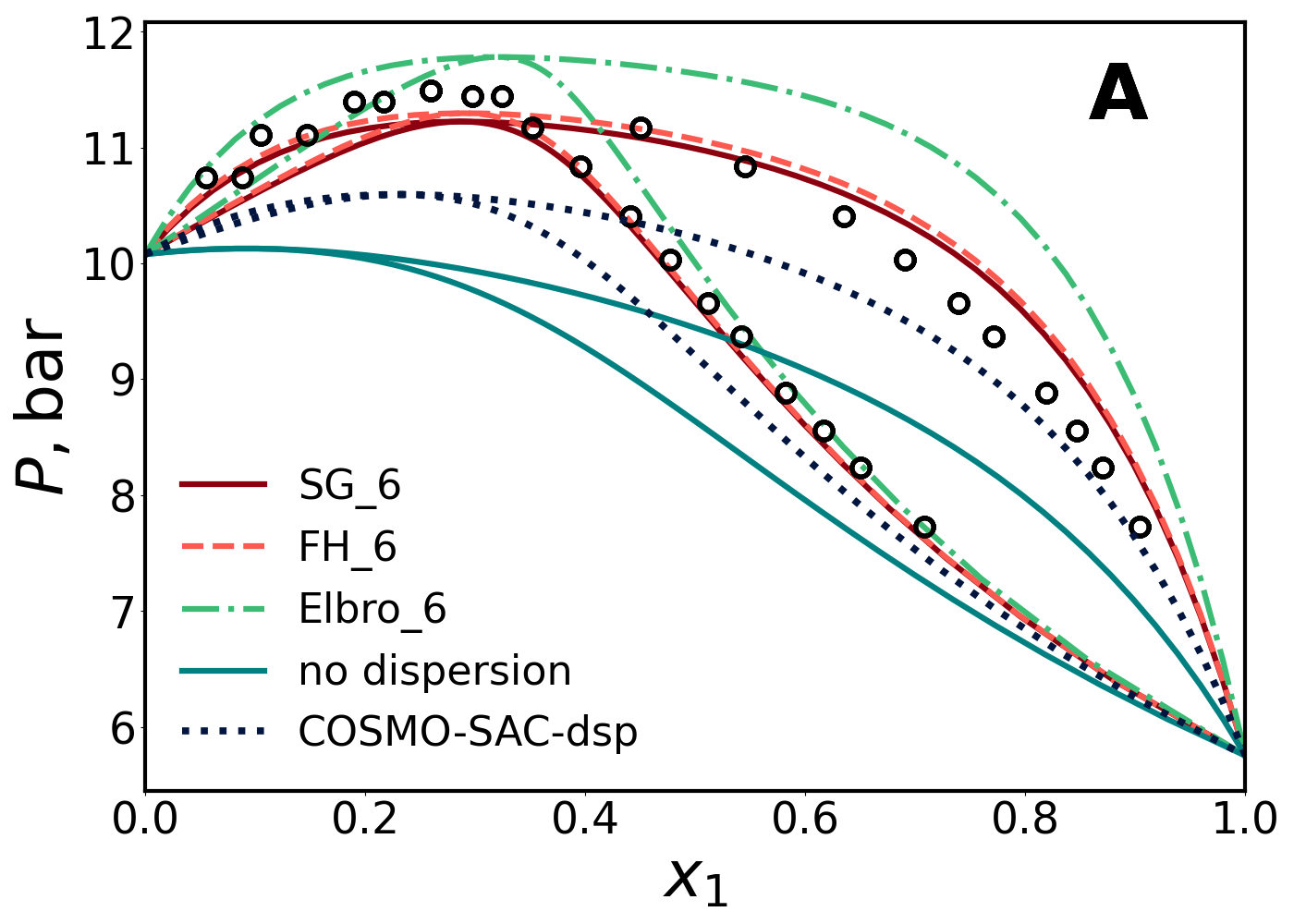}
    \includegraphics[width=0.45\linewidth]{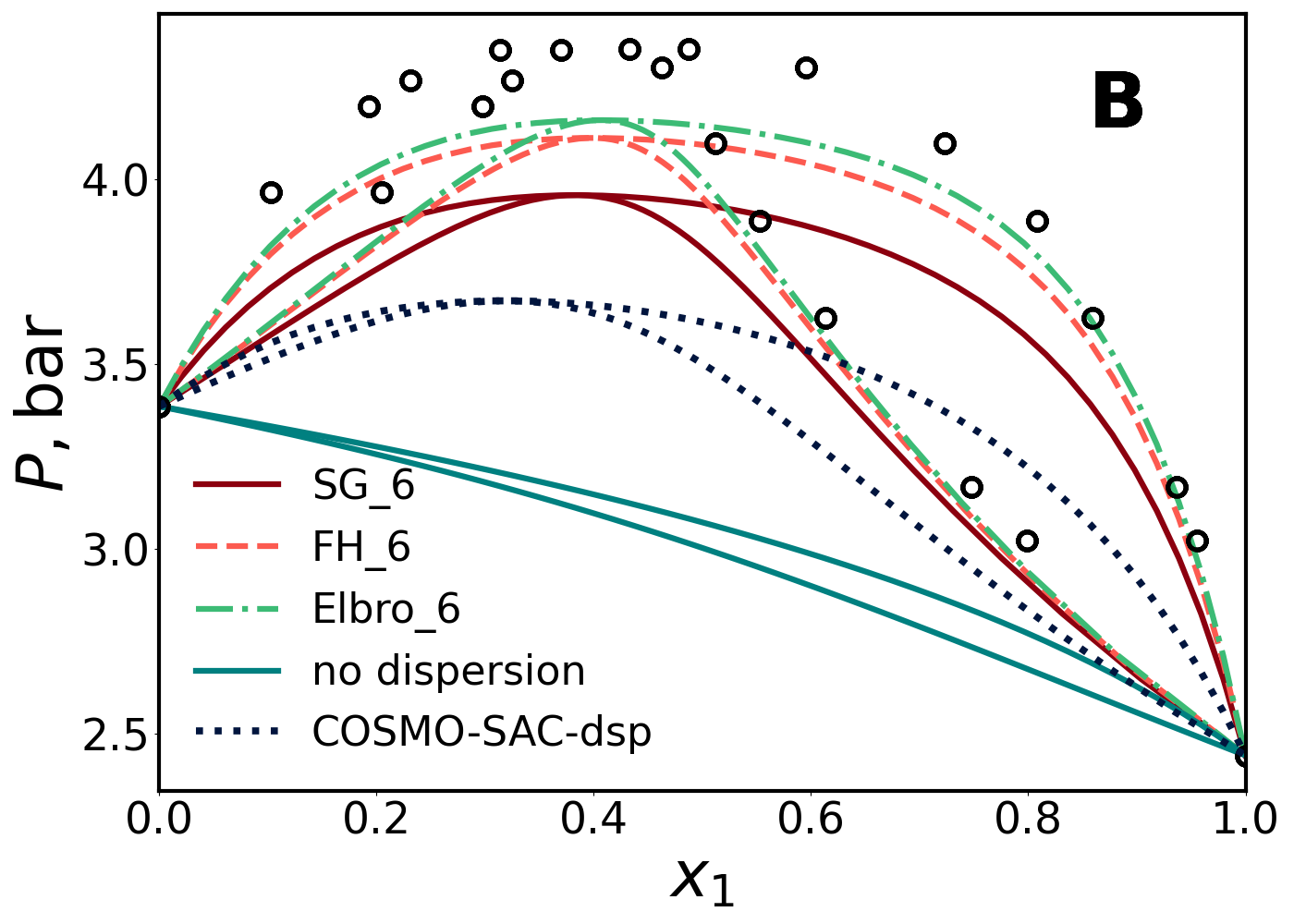}
    \caption{P-\textit{xy} phase diagram of isopentane - 1,1,1,3,3-pentafluoropropane at 362.94 K (\textbf{A}) and of propane - pentafluoroethane at 253.15 K (\textbf{B}). Experimental data \textbf{A} - \citep{ElAhmar2012VapourliquidStates} and \textbf{B} - \citep{Kim2003VaporliquidK} is represented by symbols (\textbf{o}), while corresponding lines are the predictions made using various openCOSMO-RS dispersion parametrizations and COSMO-SAC-dsp.}
    \label{fig:vleKDB4818_1}
\end{figure}

To demonstrate the impact of the dispersion term on VLE predictions, Figures \ref{fig:vleKDB4818_1}A and \ref{fig:vleKDB4818_1}B present two P-\textit{xy} phase diagrams of fluorinated refrigerant mixtures. In the absence of the dispersion contribution, the model fails to capture azeotropic behavior and even predicts nearly ideal behavior for propane - pentafluoroethane (Figure \ref{fig:vleKDB4818_1}B). However, incorporating the dispersion term, even via the simplest models not considering C-atom hybridization or cross-interactions, leads to more physically meaningful predictions. All illustrated models successfully capture the azeotrope. Moreover, the choice of combinatorial term does not have a substantial effect in those cases.

\begin{figure}
    \centering
    \includegraphics[width=0.5\linewidth]{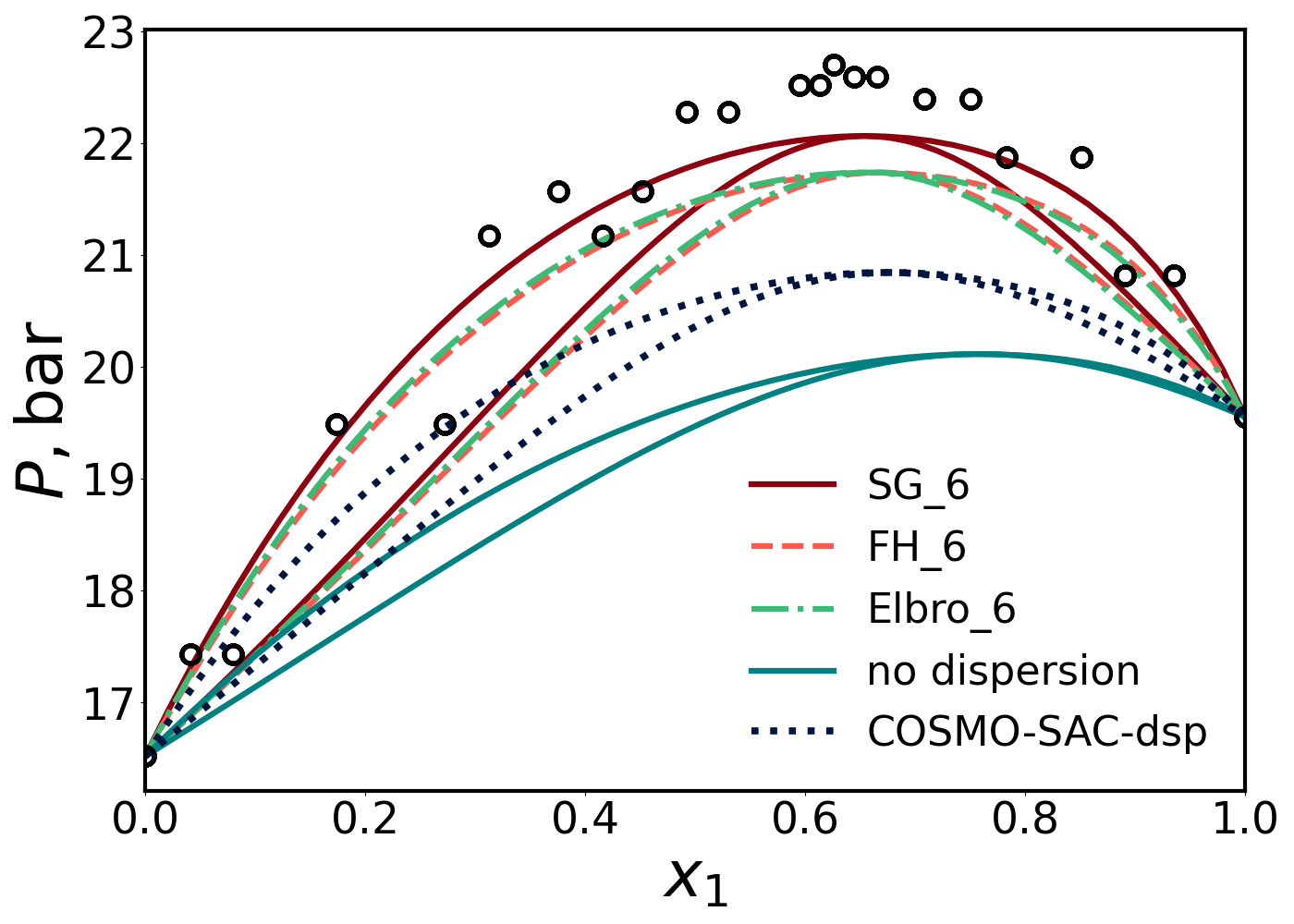}
    \caption{P-\textit{xy} phase diagram of chlorodifluoromethane - octafluoropropane at 323.15 K. Experimental data \citep{CHERIC2024} is represented by symbols (\textbf{o}), while corresponding lines are the predictions made using various openCOSMO-RS dispersion parametrizations and COSMO-SAC-dsp.}
    \label{fig:3968_IsothermalData_CHLORODIFLUROMET_OCTAFLUROPROP_at_50.0C}
\end{figure}

The maximum saturation pressure for the selected VLE systems is approximately 20 bar. At higher pressures, one should consider the non-idealities in the gas phase. Moreover, for certain mixtures, even at 20 bar, the non-ideal gas behaviour could not be ignored. However, as depicted in Figure \ref{fig:3968_IsothermalData_CHLORODIFLUROMET_OCTAFLUROPROP_at_50.0C}, all dispersion models without cross-interactions are in a good agreement with the experimental data, although some points are at pressures higher than 20 bar.

\begin{figure}
    \centering
    \includegraphics[width=0.45\linewidth]{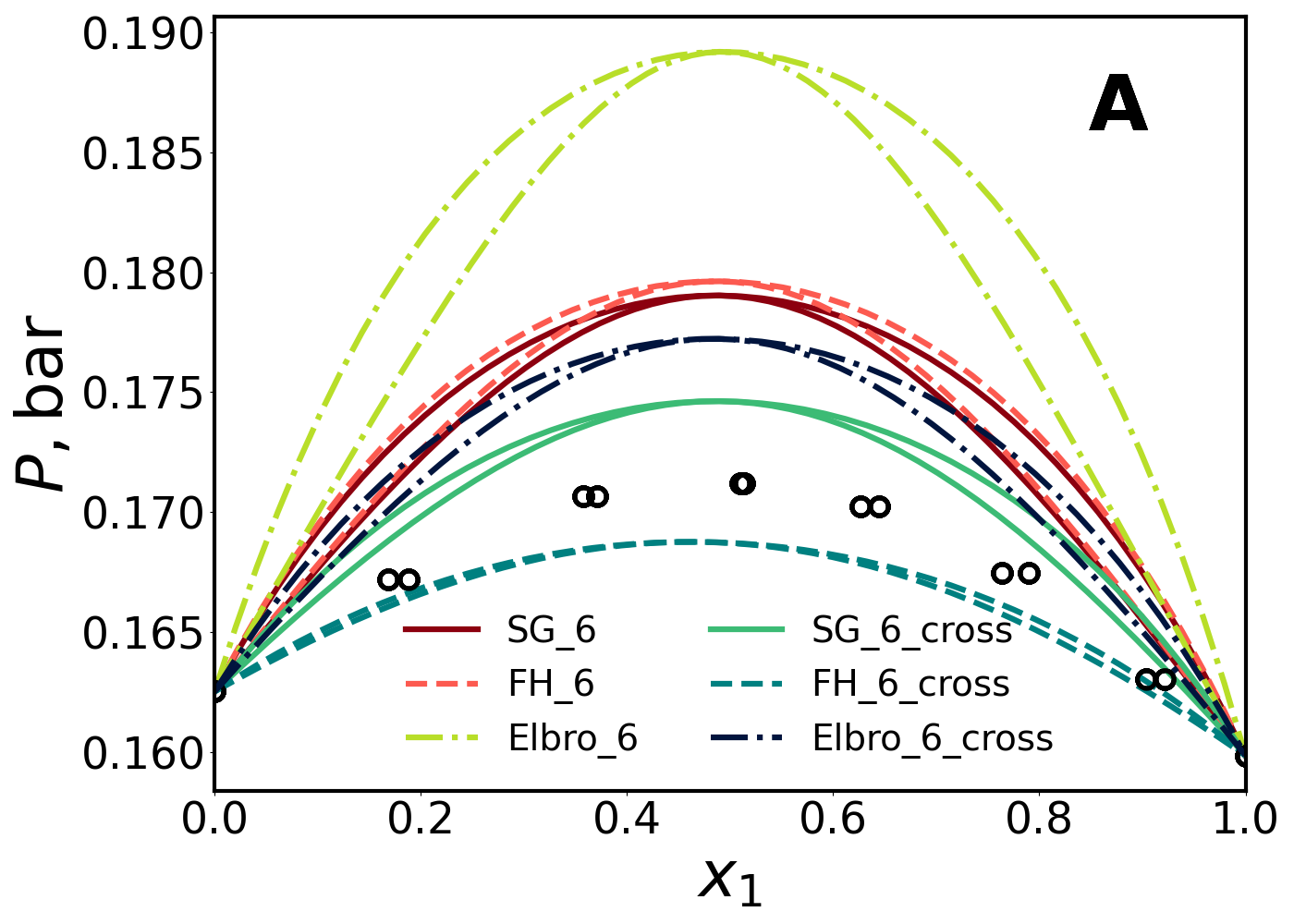}
    \includegraphics[width=0.45\linewidth]{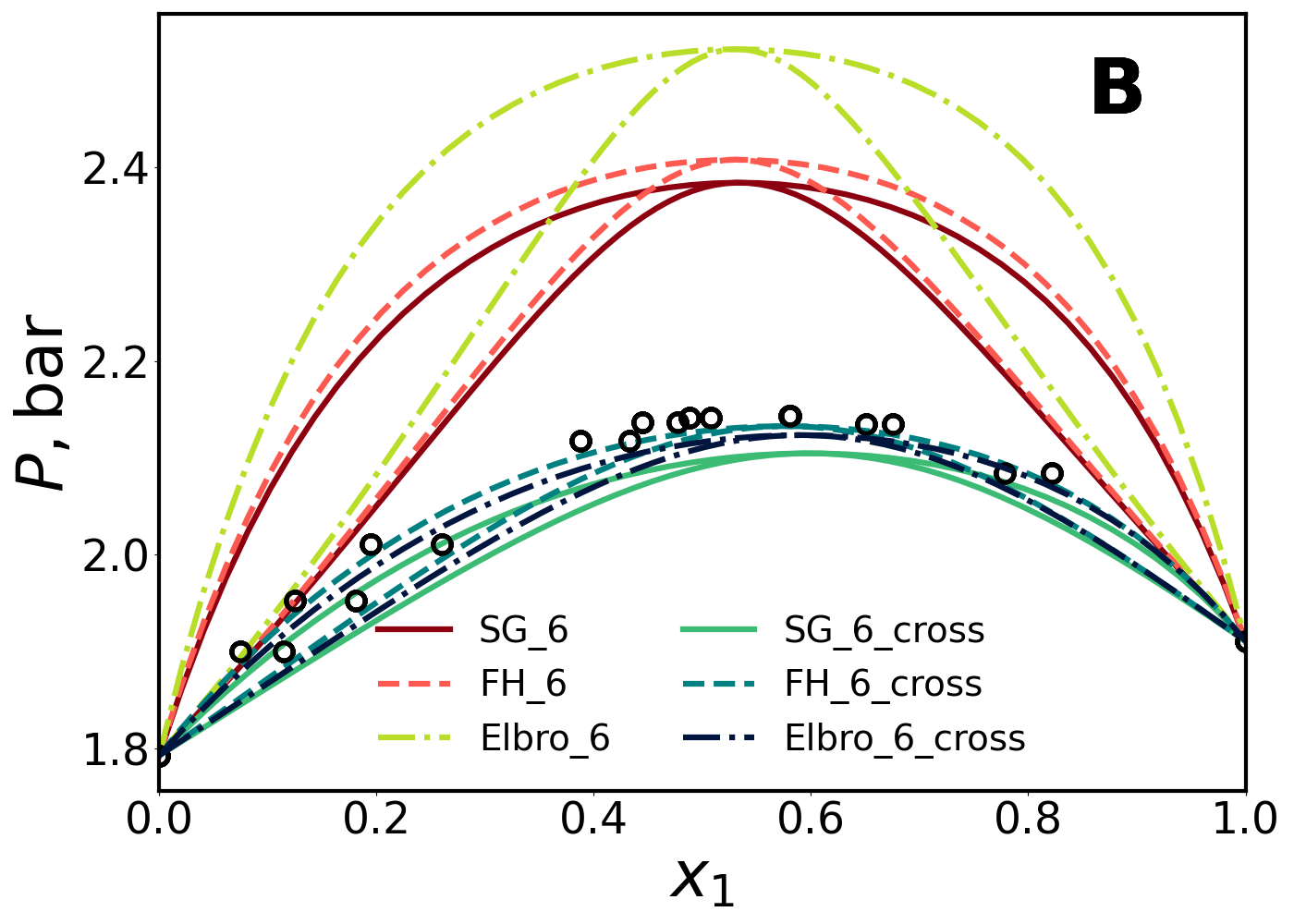}
    \caption{P-\textit{xy} phase diagrams of 1,2-dibromoethane - chlorobenzene at 348.12 K (\textbf{A}) and of trifluoroiodomethane - trans-1,3,3,3-tetrafluoropropene at 268.15 K (\textbf{B}). Experimental data \textbf{A} - \citep{Lacher1941} and \textbf{B} - \citep{Guo2012Vapour+liquid298.150K} is represented by symbols (\textbf{o}), while corresponding lines depict predictions made using various openCOSMO-RS dispersion parametrizations.}
    \label{fig:Sheet37}
\end{figure}

\begin{figure}
    \centering
    \includegraphics[width=1\linewidth]{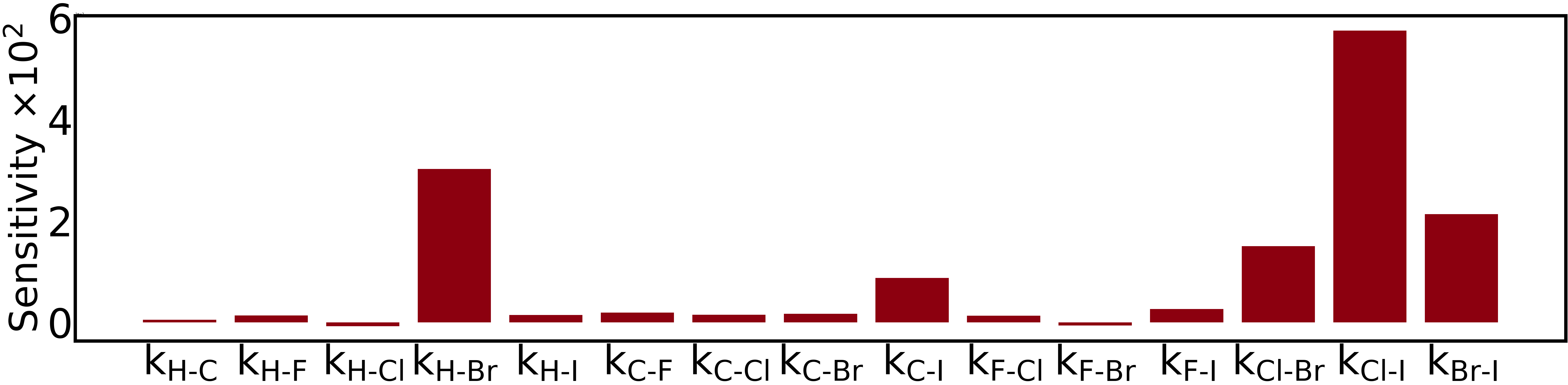}
    \caption{Local sensitivities for cross-atom parameters of SG\_6\_cross model estimated for VLE data.}
    \label{fig:sensitivity_VLE_SG_6_cross}
\end{figure}

As reported in Tables \ref{tab:param_deviation} and \ref{tab:param_deviation_cross}, the inclusion of additional cross-interaction parameters improves predictions for various types of phase equilibrium but not significantly. In the case of VLE, most systems are modeled similarly with and without additional cross-interaction parameters \(k_{ij}\). Nevertheless, certain mixtures containing multiple halogen atoms benefit from the additional \(k_{ij}\). For instance, as shown in Figures \ref{fig:Sheet37}A and \ref{fig:Sheet37}B, the models lacking cross-interaction parameters tend to overestimate saturation pressures. To better understand the impact of the cross-interaction parameters, we estimated the local sensitivities of the \(k_{ij}\) using the backward difference method mentioned previously (Eq.\ref{eq:backward_sensitivity}). One example is Figure \ref{fig:sensitivity_VLE_SG_6_cross} that demonstrates the local sensitivity analysis for the SG\_6\_cross parametrization. Notably, the highest sensitivities are mostly attributed to the halogen-halogen parameters (e.g. Cl-I, Br-I, Cl-Br, F-I), which are responsible for the improved predictions in the discussed examples. Although there is no clearly defined trend of the \(k_{ij}\) parameters, one could notice some correlation between the sign and value of \(k_{ij}\) and the size differences of the interacting atoms. For instance, accounting for \(k_{ij}\) typically increases \( E_{vdW}\) compared to \(k_{ij}\) = 0 for interactions involving the I-atom, while the opposite is generally true for interactions with H-atom.

\begin{figure}
    \centering
    \includegraphics[width=0.45\linewidth]{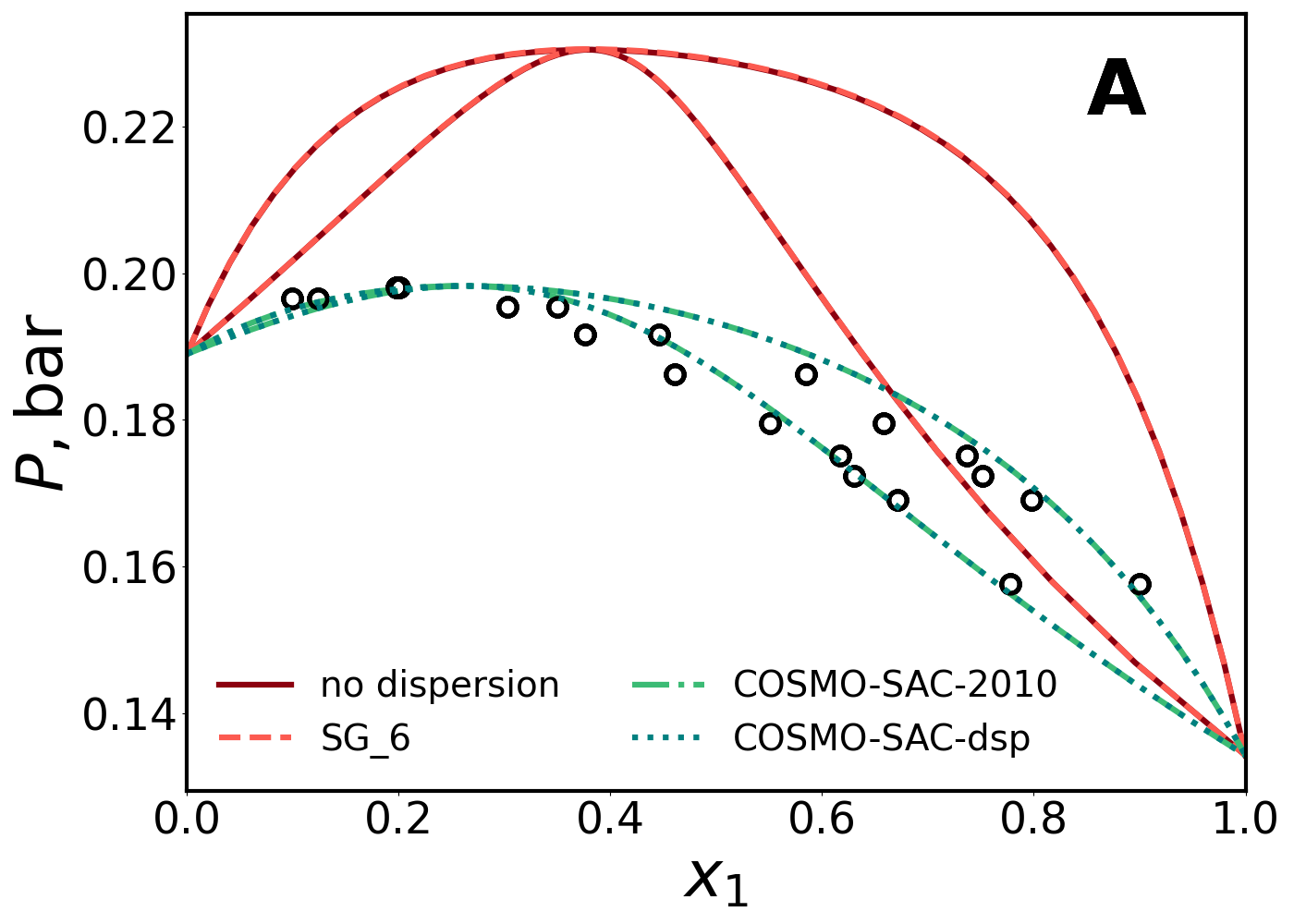}
    \includegraphics[width=0.45\linewidth]{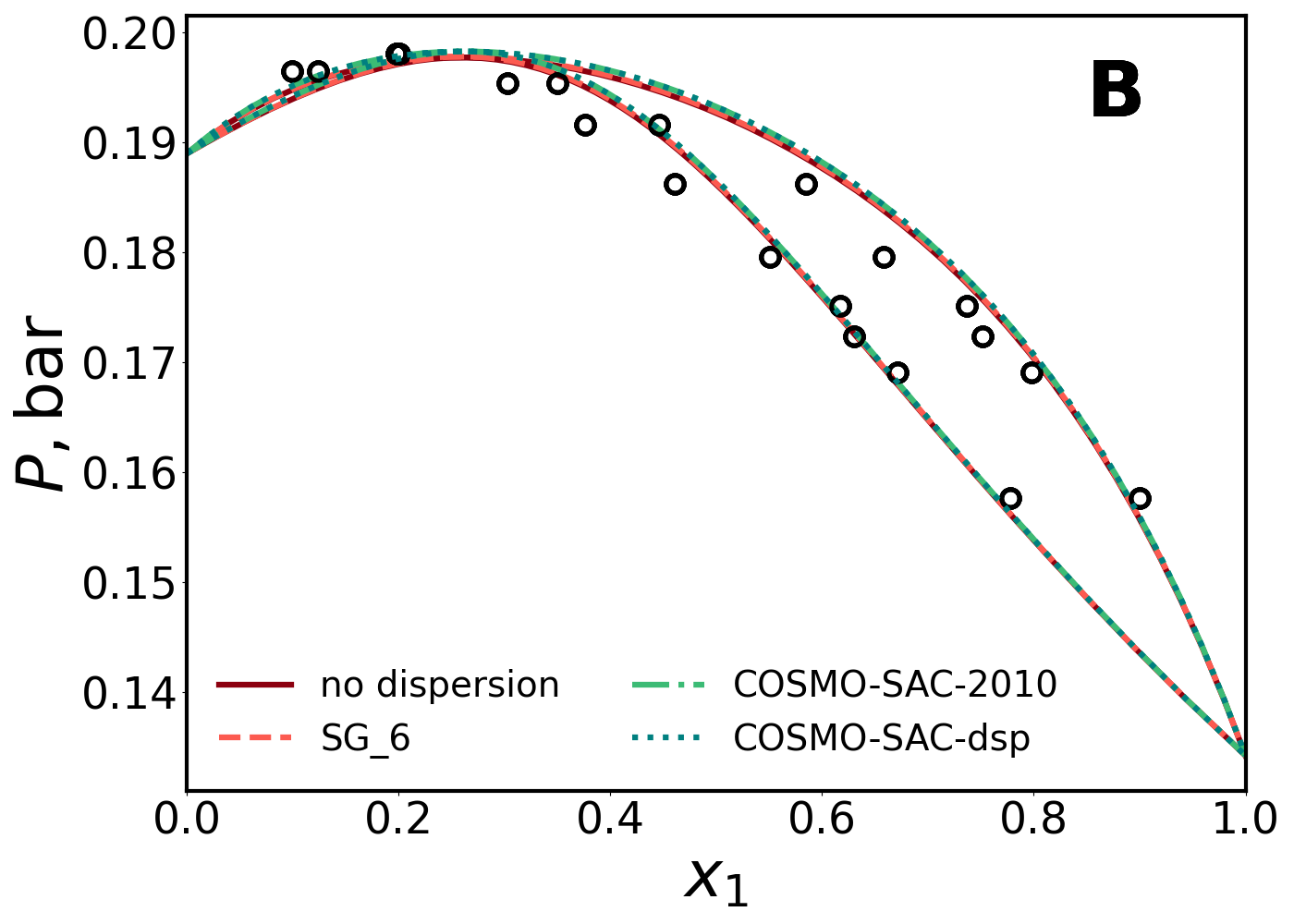}
    \caption{P-\textit{xy} phase diagrams of 1,2-dichloroethane - tetrachloromethane at 303.15 K. Experimental data \citep{Jaubert2020BenchmarkAccuracy} is represented by symbols (\textbf{o}), while corresponding lines depict predictions made using various openCOSMO-RS and COSMO-SAC models. For openCOSMO-RS, the \textit{cis}-conformer of 1,2-dichloroethane is used in (\textbf{A}) and\textit{ trans}-conformer is used in (\textbf{B}).}
    \label{fig:conformer_case}
\end{figure}

The workflow utilized for the QC calculations had some difficulty in identifying the most stable conformer for two molecules: 1,2-dichloroethane and 1,2-dibromoethane. For instance, we observed unreasonably high deviations between predictions made using openCOSMO-RS and COSMO-SAC-dsp for the 1,2-dichloroethane - tetrachloromethane mixture, irrespective of the dispersion term used. These significant discrepancies arise due to the utilization of different conformers in calculations. Initially (Figure \ref{fig:conformer_case}A), the workflow determined the \textit{cis}-conformer of 1,2-dichloroethane as the most stable. However, it is widely acknowledged that \textit{trans}-conformers are typically more stable. Consequently, we repeated the calculations using the more stable \textit{trans}-conformer (Figure \ref{fig:conformer_case}B), resulting in a remarkable difference. Although the assessment of conformers is beyond the scope of the present study, it is essential to emphasize their critical role in COSMO-RS predictions \citep{KLAMT2005109}. Therefore, an improved conformer search algorithm is currently under development.

\begin{figure}
    \centering
    \includegraphics[width=0.5\linewidth]{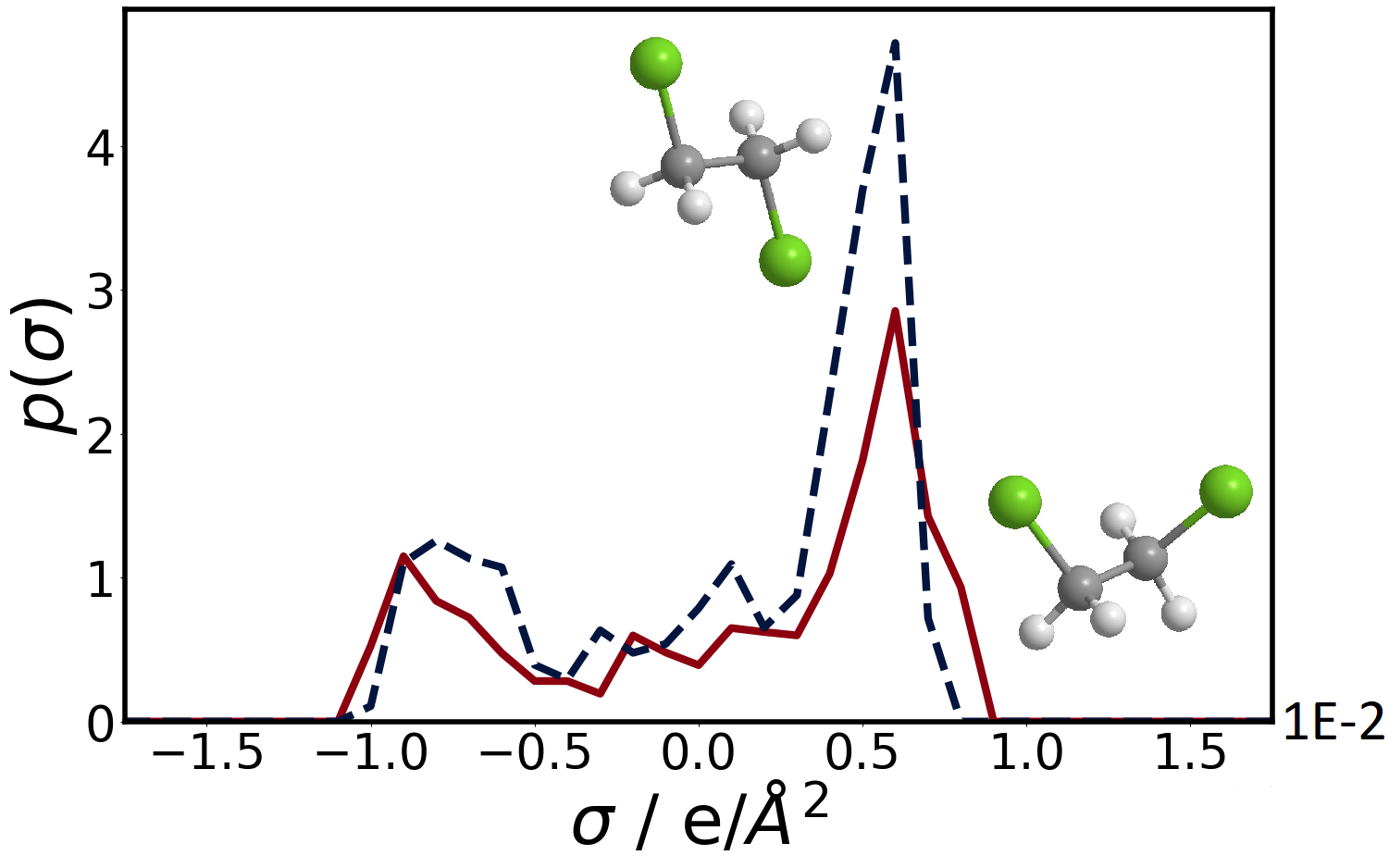}
    \caption{\(\sigma\)-profiles of \textit{cis}- and \textit{trans}- conformers of 1,2-dichloroethane. The 3D structure of the conformers is visualized using Chem3D software.}
    \label{fig:sigma_profiles_12dichloroethane}
\end{figure}

\begin{figure}
    \centering
    \includegraphics[width=0.5\linewidth]{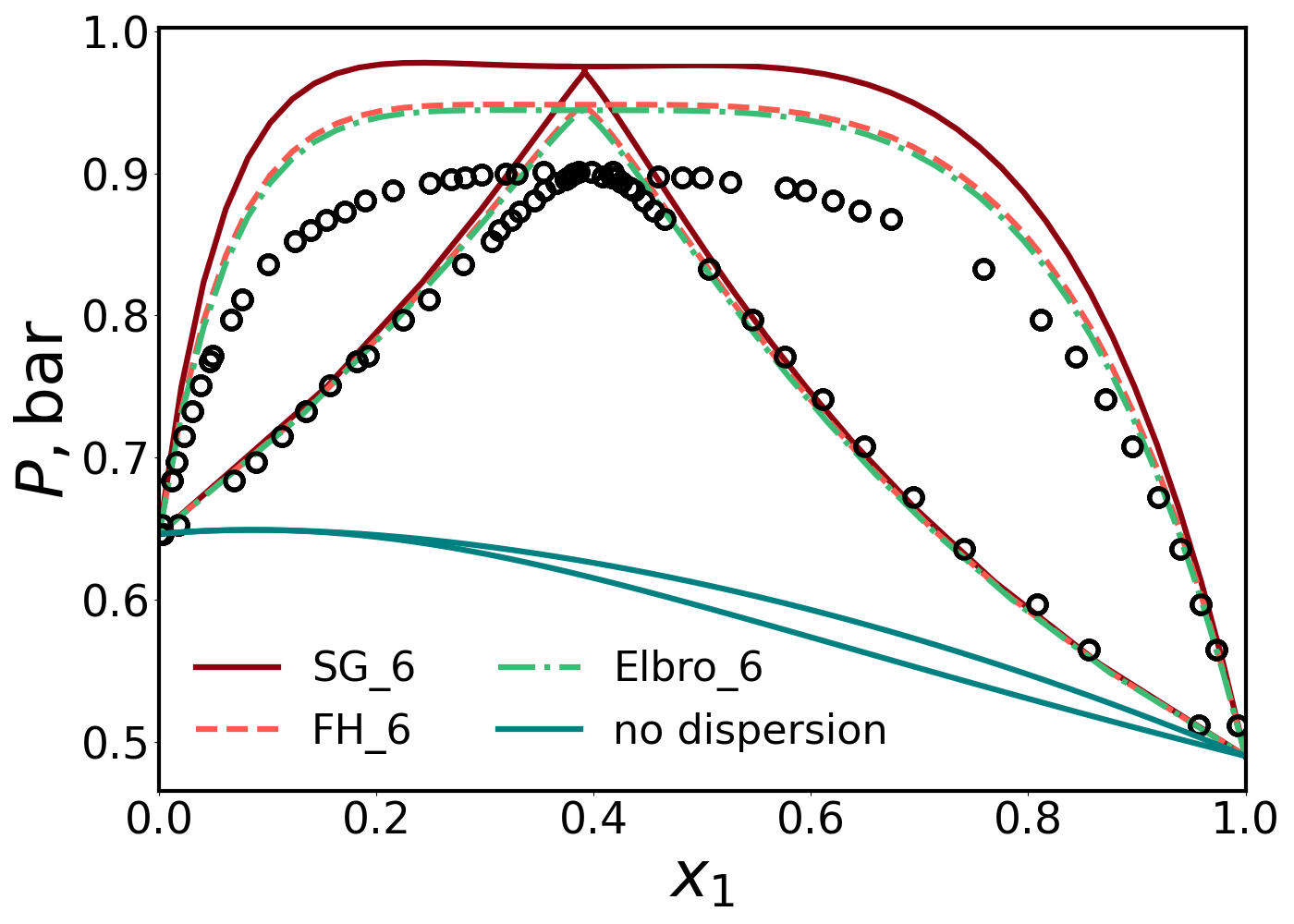}
    \caption{P-\textit{xy} phase diagram of perfluoromethylcyclohexane - n-hexane at 328.15 K. Experimental data \citep{CHERIC2024} is represented by symbols (\textbf{o}), while corresponding lines depict predictions made using various openCOSMO-RS models.}
    \label{fig:perfluoromethylcyclohexane_fail}
\end{figure}

Despite the significant improvements in VLE predictions due to incorporation of the dispersion term into openCOSMO-RS, several challenges remain. One such challenge is illustrated in Figure \ref{fig:perfluoromethylcyclohexane_fail}, where poor predictions may arise from neglecting the steric hindrance and the equilibrium distribution of conformers. The model without the dispersion contribution predicts a nearly ideal mixture, while other dispersion models overestimate the saturation pressures. This discrepancy could be attributed to the effect of steric hindrance.   

\begin{figure}
    \centering
     \includegraphics[width=0.5\linewidth]{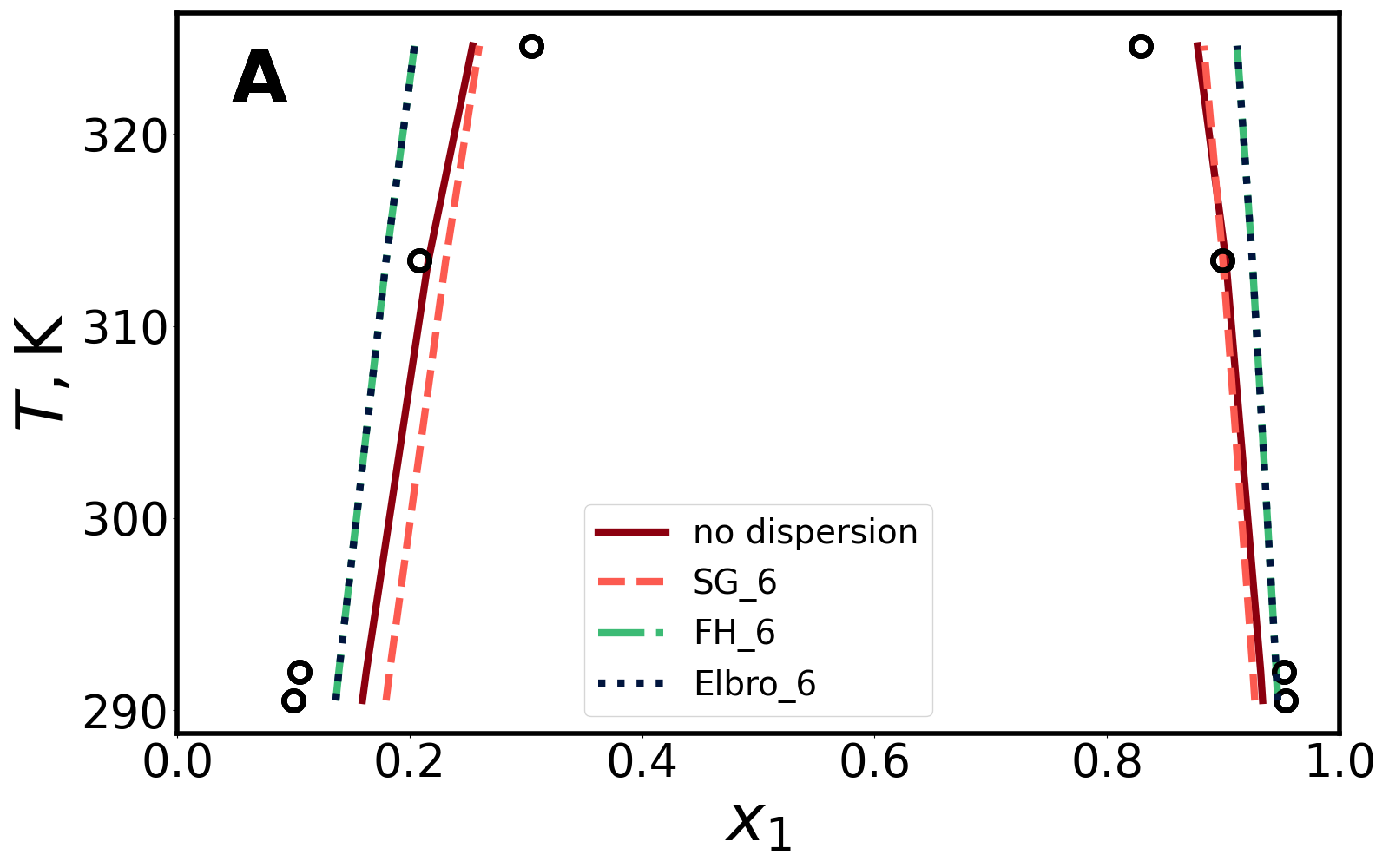}
     \includegraphics[width=0.5\linewidth]{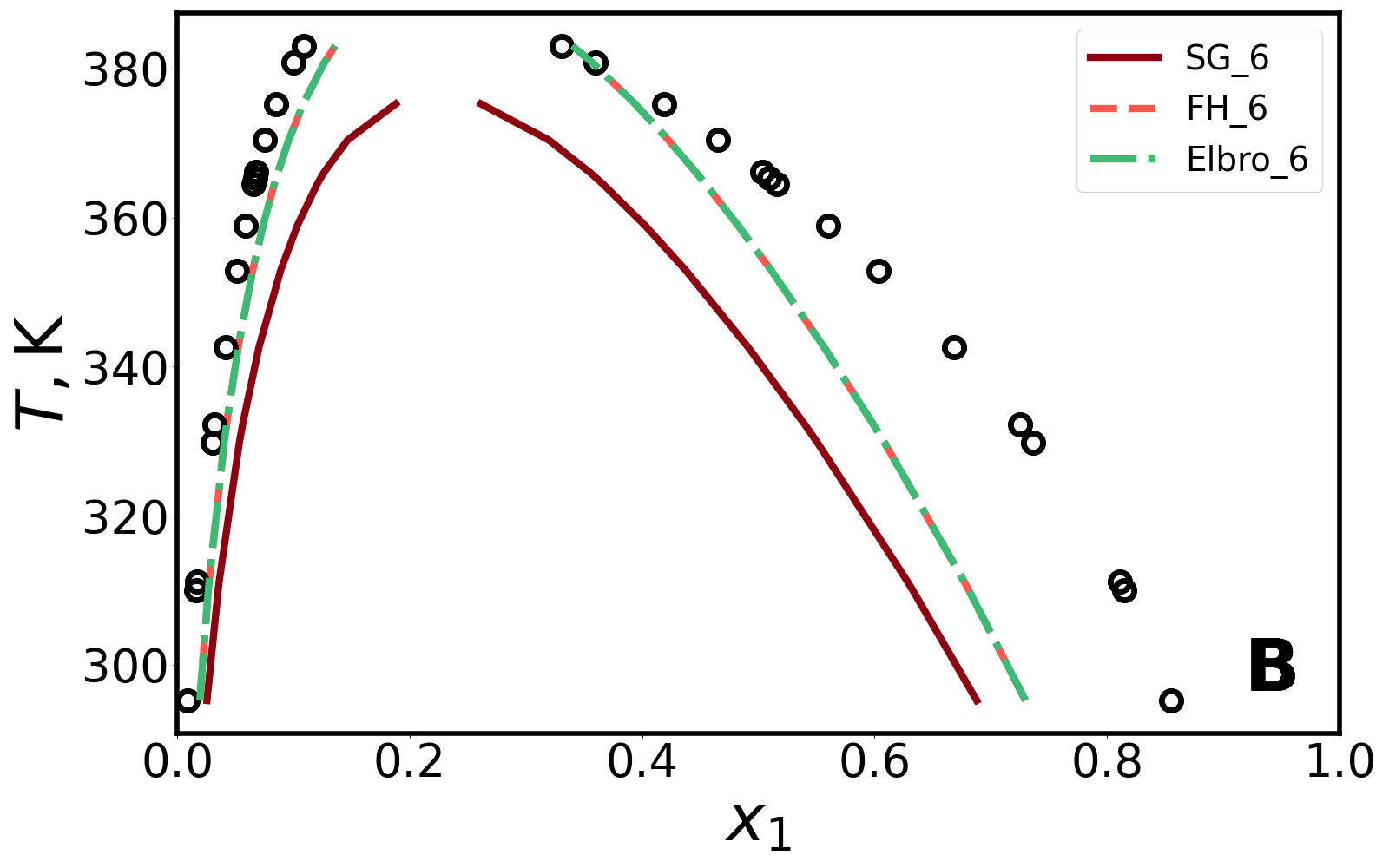}
     \includegraphics[width=0.5\linewidth]{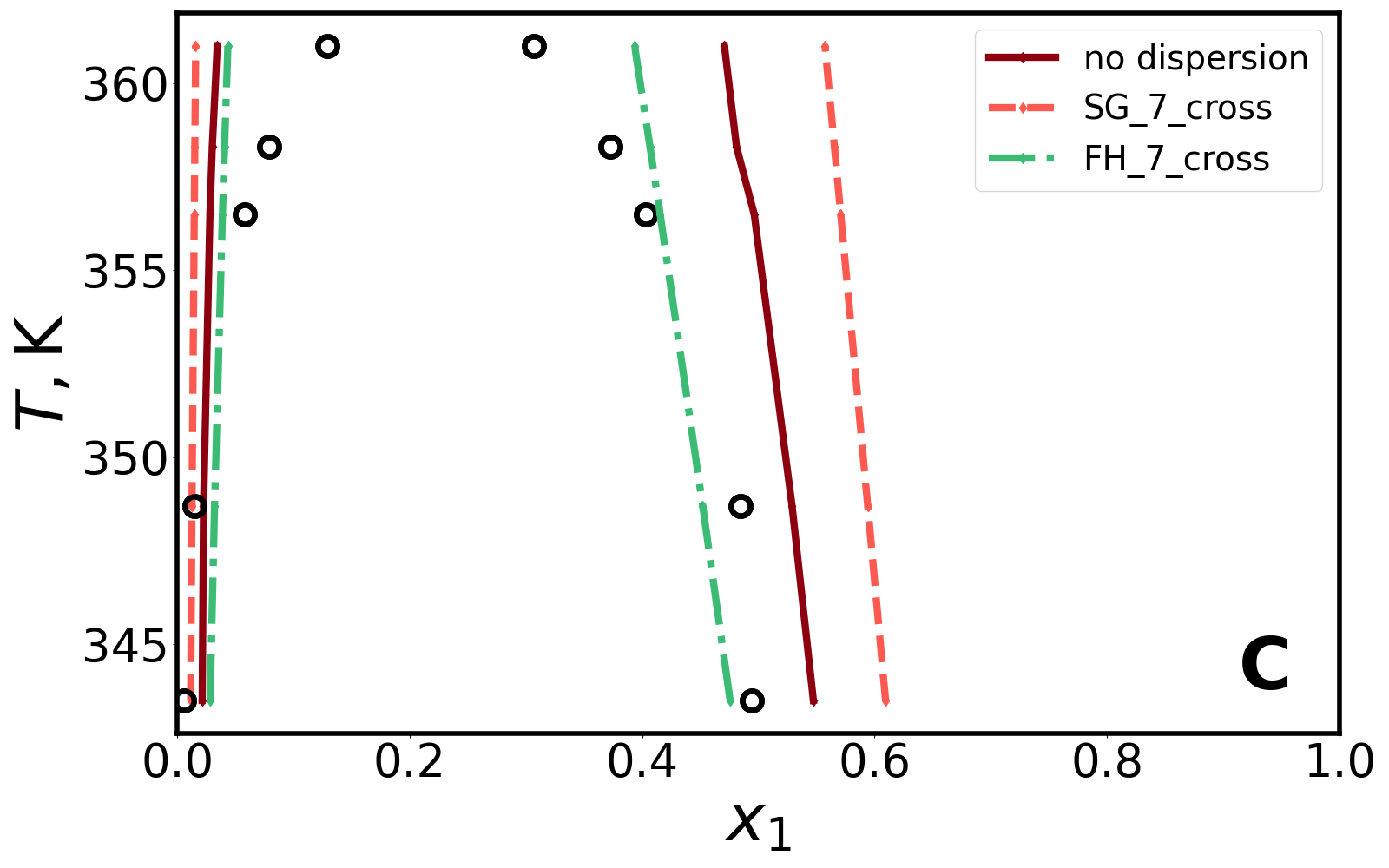}
    \caption{LLE of the octane - perfluoroheptane (\textbf{A}), the perfluoroheptane - benzene (\textbf{B}), and the octadecafluorooctane - dichloromethane (\textbf{C}) systems. Experimental data \citep{LLE10} is represented by symbols (\textbf{o}), while corresponding lines depict predictions made using various openCOSMO-RS models.}
    \label{fig:LLE_octane_hexadecafluoro-n-heptane}
\end{figure}

In LLE predictions, both openCOSMO-RS without the dispersion term and COSMO-SAC-dsp fail to identify any phase split for more than half of the collected systems. However, in certain cases, such as Figures \ref{fig:LLE_octane_hexadecafluoro-n-heptane}A and \ref{fig:LLE_octane_hexadecafluoro-n-heptane}C, reasonable predictions were achieved regardless of the absence of the dispersion term. As demonstrated in Tables \ref{tab:param_deviation} and \ref{tab:param_deviation_cross}, models utilizing the SG combinatorial term are slightly inferior to those using the FH or Elbro terms. Specifically, models involving the FH combinatorial term lead to the lowest deviations between experimental data and predictions for LLE. For instance, in Figure \ref{fig:LLE_octane_hexadecafluoro-n-heptane}B, the SG\_6 parametrization captures the curvature of the phase split. However it underestimates the critical solution temperature, which is a notoriously challenging property to predict. Conversely, predictions by FH\_6 and Elbro\_6 are in better agreement with the experimental data. It is noteworthy that capturing the dome - shape of the binodal curve is a general challenge for all parametrizations across the majority of systems, regardless of the number of parameters. One can observe in Figure \ref{fig:LLE_octane_hexadecafluoro-n-heptane}C that even the most accurate parametrization for LLE FH\_7\_cross does not entirtely follow the dome - shape of the binodal curve. 

The impact of the combinatorial term on VLE and LLE is not as pronounced as on IDAC data, although still substantial. Despite our expectations after analyzing IDAC predictions, the Elbro term did not demonstrate an outstanding performance for other types of phase equilibrium evaluated in this study. AAD values for VLE data listed in Tables \ref{tab:param_deviation} and \ref{tab:param_deviation_cross} indicate its less satisfactory performance compared to other models. However, for LLE data, the Elbro term performs almost as well as the FH term. One could notice from the list of systems in Tables S1 and S2 of the Supplementary Materials that LLE datasets comprise more asymmetrical systems than VLE datasets. The advantage of the Elbro term in modeling asymmetric systems agrees with previous findings. The SG term models tend to produce VLE predictions the most accurately. Although the FH term leads to the poorest predictions of IDACs, it seems to be the best choice for LLE. Therefore, across all the parametrizations obtained in this study, there are no universal solutions leading to the superior predictions for all types of phase equilibrium. Nevertheless, when modeling halocarbons/refrigerants phase equilibrium, we consider the FH term to be the optimal choice. Models incorporating the FH term predict all types of the phase equilibrium reasonably well. Unlike the SG term, it does not require an additional general parameter to the model, and unlike Elbro, it does not require density experimental data. That said, in general, the performance of the combinatorial terms varies for different phase equilibrium types, however in the light of developing a general predictive model, finding a satisfactory compromise is important. In the future work, we plan to extend our analysis on a larger dataset with more chemically diverse compounds.

\section{Conclusions}
\label{sec:conclusions}
In this study, we improved openCOSMO-RS by incorporating the dispersion contribution and applied the modified version to predict phase equilibrium of halocarbons and refrigerants. We collected and evaluated an extensive database of VLE, LLE, and IDAC to parameterize and assess the model's performance. 

Furthermore, our analysis showed that differentiating between carbon atom hybridizations did not yield significant improvements in predictions. While extending the parameter set with \(k_{ij}\) cross-atom parameters proved beneficial for certain systems with multiple halogens, the overall improvements were not substantial compared to parametrizations without them. Therefore, a relatively simple parameter set without additional \(k_{ij}\) and without a second parameter for carbon atom proved sufficient for satisfactory phase equilibrium predictions of the considered mixtures.

Additionally, we evaluated the performance of several combinatorial terms. The theoretically well-supported Elbro term demonstrated remarkable accuracy in predicting IDAC data of alkanes. However, for VLE, its accuracy was inferior to the SG and FH terms. Specifically, the FH term performed better for LLE, while the SG term excelled for VLE.

Overall, for modeling VLE and LLE of halocarbons and refrigerants VLE and LLE, we recommend using the FH term coupled with the dispersion parametrization consisting of six atom-specific parameters. In future research, we aim to implement the dispersion contribution for a broader range of systems.


\bibliographystyle{elsarticle-harv} 
\bibliography{CES_ref_template}




\end{document}